\documentclass{aa}  

\usepackage{graphicx}
\usepackage{txfonts}
\usepackage{siunitx}
\usepackage{natbib,twoopt}
\usepackage{xcolor}
\usepackage{ulem}
\usepackage[breaklinks=true]{hyperref} %% to avoid \citeads line fills
\bibpunct{(}{)}{;}{a}{}{,}             %% natbib format for A&A and ApJ
\makeatletter
  \newcommandtwoopt{\citeads}[3][][]{\href{http://adsabs.harvard.edu/abs/#3}%
    {\def\hyper@linkstart##1##2{}%
     \let\hyper@linkend\@empty\citealp[#1][#2]{#3}}}
  \newcommandtwoopt{\citepads}[3][][]{\href{http://adsabs.harvard.edu/abs/#3}%
    {\def\hyper@linkstart##1##2{}%
     \let\hyper@linkend\@empty\citep[#1][#2]{#3}}}
  \newcommandtwoopt{\citetads}[3][][]{\href{http://adsabs.harvard.edu/abs/#3}%
    {\def\hyper@linkstart##1##2{}%
     \let\hyper@linkend\@empty\citet[#1][#2]{#3}}}
  \newcommandtwoopt{\citeyearads}[3][][]%
    {\href{http://adsabs.harvard.edu/abs/#3}
    {\def\hyper@linkstart##1##2{}%
     \let\hyper@linkend\@empty\citeyear[#1][#2]{#3}}}
\makeatother

\begin{document}

   \title{Small-scale loops heated to transition region temperatures and their chromospheric signatures in the simulated solar atmosphere}

   %\subtitle{I. Overviewing the $\kappa$-mechanism}

   \author{M. Skan \inst{1}
          \and
          S. Danilovic \inst{1}
          \and
          J. Leenaarts \inst{1}
          \and
          F. Calvo \inst{1}
          \and
          M. Rempel \inst{2} 
          %\fnmsep\thanks{Just to show the usage
          %of the elements in the author field}
          }

   \institute{Institute for Solar Physics,
                Department of Astronomy,
                Stockholm University,
              Roslagstullsbacken 21, Stockholm\\
              \email{moa.skan@astro.su.se}
         \and High Altitude Observatory, National Center for Atmospheric Research, 80307, Boulder, CO, USA \\
             }
 \titlerunning {Small-scale loops heated to TR temperatures}
 \authorrunning {Skan et al.}
   \date{}

% \abstract{}{}{}{}{} 
% 5 {} token are mandatory
 
  \abstract
   {Recent observations revealed loop-like structures at very small scales visible in observables that sample transition region (TR) and even coronal temperatures. Their formation remains unclear.}
   {Realistic magnetohydrodynamic simulations and forward synthesis of spectral lines are used to investigate how these features occur.}
   {Computations are done using the MURaM code to generate model atmospheres. The synthetic H$\alpha$ and Si~IV spectra are calculated at two angles ($\mu=1$, $\mu=0.66$) using the Multi3D code. Magnetic field lines are traced in the model and the evolution of the underlying field topology is examined.}
   {The synthetic H$\alpha$ dopplergrams reveal loops that evolve dramatically within a few minutes. The synthetic H$\alpha$ line profiles show observed asymmetries and doppler shifts in the line core. They, however, also show strong emission peaks in the line wings, even at the slated view. The synthetic Si~IV emission features partly coincide with structures visible in H$\alpha$ dopplergrams and partly follow separate magnetic field threads. Some are even visible in the emission measure maps for the lg($T$ /K)$ = [5.0 , 5.5]$ temperature interval. The emission areas trace out the magnetic field lines rooted in opposite polarities in a bipolar region.}
   {The model largely reproduce the observed features namely their size, lifetime and morphology in both observables. A bipolar system with footpoints undergoing rapid movement and shuffling can produce many small-scale recurrent events heated to high temperatures. The morphology and evolution of the resulting observable features can vary depending on the viewing angle.}


   \maketitle
%
%-------------------------------------------------------------------

\section{Introduction}
The coronal heating problem has been a widely discussed subject since it was discovered roughly eight decades ago when \citet{1939NW.....27..214G} saw emission lines in the corona that suggested that the coronal plasma is at very high temperatures. It is widely accepted that the motions of the plasma in the convective zone is a big source of the energy injected into the corona, but how the energy is transported and dissipated is a standing question. The two main theories that have been accepted by the community are \textit{direct current} (DC) and \textit{alternating current} (AC) heating (\citet{2006SoPh..234...41K}) which differ in relevant time scales on which convective motions affect the footpoints of magnetic field lines. The mechanism behind DC heating is the slow gradual buildup of stresses due to the displacement of the footpoints of magnetic field lines. As a consequence, the magnetic field higher up in the atmosphere is being contracted, expanded, or twisted to some degree and pushed together which results in the creation of current sheets. The current sheets dissipate gently through Ohmic heating and nanojets (\citet{1988ApJ...330..474P}), or violently by magnetic reconnection. Several numerical experiments have been done on the subject, see e.g.\ \citep{2005ApJ...618.1020G, 2013A&A...550A..30B, 2015ApJ...811..106H, 2017ApJ...834...10R}. In these models, heating is highly intermittent in both space and time. The heating structures are loop-like as they follow the magnetic field lines. Their lifetimes vary from less than a minute to an hour, depending on their size and the unsigned magnetic flux in the model.\\
\newline
Over the last decade, the theoretical models got further challenged by the new instrumentation that provides observations of the solar atmosphere in increasingly finer detail. The \textit{High-resolution Coronal Imager} \citep[Hi-C,][]{2019SoPh..294..174R} showed features as small as $1-5$~Mm in length apparently heated to coronal temperatures ($>1$~MK) \citep{2013A&A...556A.104P}. Further study \citep{2017A&A...599A.137B} considered different scenarios to explain their appearance and settled on the conclusion that they might be miniature loops. It remained unclear what drives 
them to reach these extreme temperatures. \\
\newline
Interface Region Imaging Spectrograph (IRIS, \citet{2014SoPh..289.2733D}) allowed to resolve \textit{transition region} (TR) heights where the “unresolved fine structure” (UFS) loops, resided as suggested by \citet{1983ApJ...275..367F}. \citet{2014Sci...346E.315H} confirmed that these are indeed shaped out of a plethora of loop-like structures disconnected from the corona. Along with the loop-like structures on larger scales, small emission structures that reach only a few Mm in height are also found to be heated to transition region temperatures  ($4\times10^4$-$2.2\times10^5$ K).  Subsequently, \citet{2016ApJ...826L..18B} examined 109 cases of these small features and concluded that they might well be individual structures and not made up of multiple magnetic threads.  \citet{2018A&A...611L...6P} found that some of these features have counterparts in chromospheric observables sampled with the 1-m Swedish Solar Telescope (SST, \citet{2003SPIE.4853..341S}). The appearance of inverse Y-shaped structures in H$\alpha$ line wings visible above the Si~IV loops suggests that magnetic reconnection is taking place in these cases. As the authors state, this is also confirmed by the apparent opposite flows, visible in H$\alpha$ Dopplergrams, that seem to trace the magnetic field lines.  They were, however, puzzled by several facts. If the magnetic reconnection is taking place, the question remains why they find no indication of heating in chromospheric observables. Also, in some cases, they find Si~IV below, or at lower heights than  H$\alpha$ signatures. That does not agree with the scenario in which cool chromospheric plasma is being pushed into higher layers where reconnection with the overlying field occurs. The questions also remain about their role in the heating of the solar atmosphere given how infrequent their appearance is. As suggested, they might be a consequence of less frequent flux emergence which may not be the cause of heating for typical loops in the solar atmosphere.\\
\newline
Finally,  Extreme Ultraviolet Imager \citep[EUI,][]{2020A&A...642A...8R} on board the Solar Orbiter \citep{2020A&A...642A...1M} revealed many features with lengths between 400 km and 4000 km heated to apparent temperature of $\approx 1$~MK \citep{2021A&A...656L...4B}. These localized brightenings visible in $17.4$~nm passband are named ‘campfires’. They appear close to the chromospheric network and last between $10$~s and $200$~s. Using the triangulation with an assumption that these features are semi-circular loops, it is estimated that their formation height is between $1000$ and $5000$~km above the photosphere. \citet{2022A&A...660A.143K} compared the location of these features with photospheric magnetograms obtained with  Polarimetric and Helioseismic Imager \citep[PHI,][]{2020A&A...642A..11S}. They found that a large majority of campfires can be projected onto a bipolar system. The authors interpret the appearance of campfires as a sign of magnetic reconnection.\\
\newline
In this paper, we examine the evolution of a bipolar system in realistic magnetohydrodynamics (MHD) simulations. The bipolar system produces localized heating events similar to the observed small-scale structures at TR and coronal temperatures. We focus on synthetic observables H$\alpha$ and Si~IV line profiles as observed by SST and IRIS, respectively. We study their formation and how the motions of the footpoints affect their morphology and evolution. We use this to illustrate how a combination of complex magnetic topology and realistic footpoint driving can lead to the appearance of small compact features as recently observed. The sections of the article are organized as follows: Section~\ref{s:model} describes the theoretical model and the computation of synthetic observables. Section~\ref{s:results} first presents synthetic observables at two viewing angles. We show line profiles and formation heights. Then we present the result of tracing magnetic field lines and their evolution. Section~\ref{s:discussion} includes a discussion of similarities and differences to observed cases. In Section~\ref{s:conclusions} we present our conclusions. \\
\begin{figure}
    \centering
    \includegraphics[width=88mm]{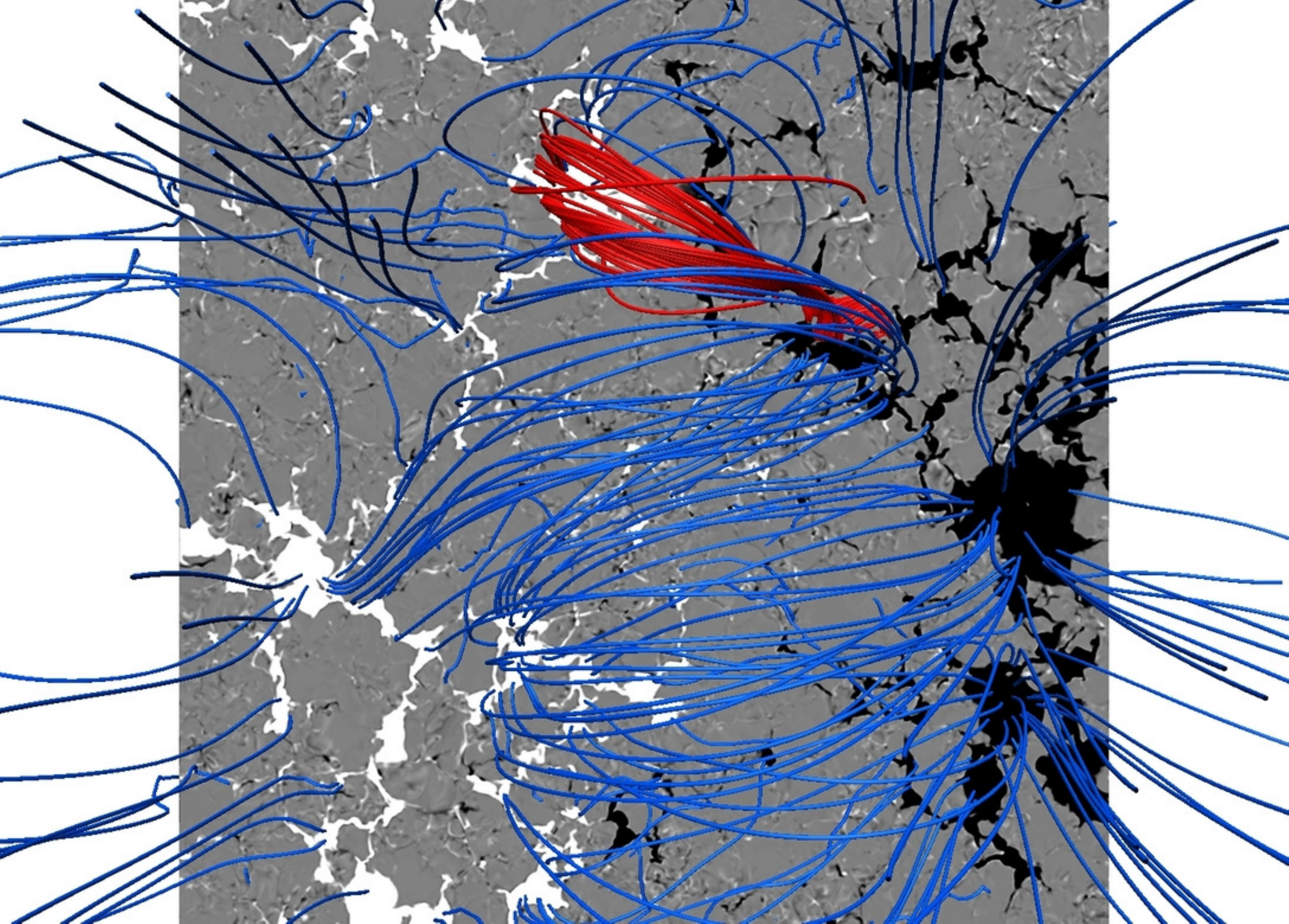}
    \caption{Overview of the full computational domain and magnetic field configuration. The plane shows the vertical photospheric magnetic field in $\mu=1$ at $t=102$~s. Red field lines show traced magnetic field lines in the region of the loop-like structure, sampled at high current density. Blue field lines show the magnetic field in the full FOV. }
    \label{fig:full_vapor}
\end{figure}  
\begin{figure*}
   \centering
   \includegraphics[width=180mm]{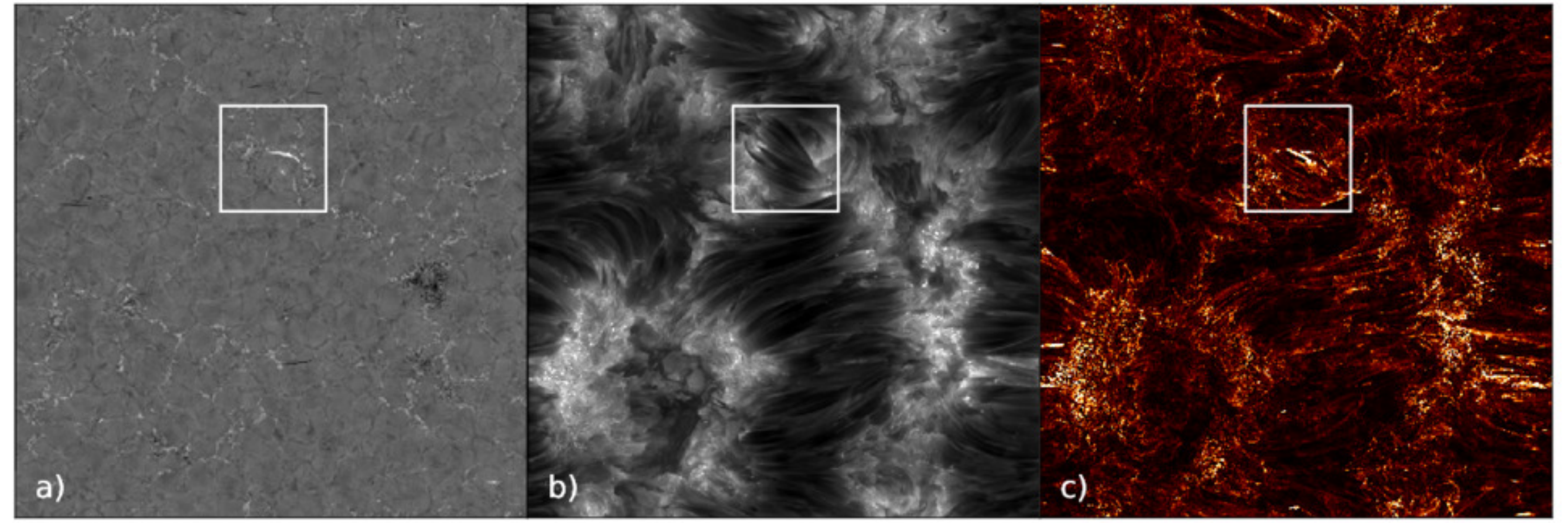}
   \caption{Synthetic observables over the full FOV at $\mu=1$, $t=102$~s. The box shows the selected subfield in figure \ref{fig:1.1}. In subplot a) we plot the blue wing of H$\alpha$, b) the H$\alpha$ line core intensity, and c) Si~IV summed over all frequencies $\pm0.79$~\AA\ around the line center.} 
    \label{fig:intro}%
\end{figure*}
%--------------------------------------------------------------------
\section{Modelling}
\label{s:model}

\subsection{Model atmospheres}
The MURaM code \citep{2005A&A...429..335V, 2017ApJ...834...10R} is used to generate the numerical 3D models of the solar atmosphere. The model includes the near-surface layers of the convection zone and self-consistently considers mass transport between subsurface layers, photosphere, chromosphere, and corona. The model is built in phases starting from the nonmagnetic convection simulation to which a uniform bipolar field of $200$~G is added. After the field configuration evolved generating magnetic structures on mesogranular and larger scales, the potential field extrapolations are used to extend the computational domain and include the upper atmosphere. The final extent of the simulation domain is $40 \times 40 \times 22$~Mm,
spanning vertically from $-8$~Mm to $14$~Mm above the
average $\tau_{500}= 1$ height. The grid spacing of the simulation used in this paper is $39$~km and $22.5$~km in the horizontal and vertical directions, respectively. The same model is used to study properties and dynamics of chromosheric fibrils \citep{2022arXiv220803744D,2022arXiv220813749D}. It contains only the most basic physics to be called ‘comprehensive’: 3D grey LTE radiative
transfer, a tabulated LTE equation of state, Spitzer heat conduction, and optically thin radiative losses in the corona based on
CHIANTI \citep{2012ApJ...744...99L}. The numerical diffusive and resistive terms in this run are set so that an effective magnetic Prandtl number Pm $ > 1$. This makes viscous heating to be the largest contributor. The time t$=0$~s is arbitrarily chosen and marks the time when the model reached the relaxed state and the snapshots are outputted every $~2.5$~s. 
Figure \ref{fig:full_vapor} displays the photospheric vertical component of the magnetic field in black/white, the magnetic field lines of the full computational domain in blue and the magnetic field lines of the \textit{region of interest} (ROI) in red. The red field lines are traced from regions with a high current density in the area.

% Realistic microphysics assumptions. Numerical resistivity.

\subsection{Radiative transfer computations}
Multi3D (\citet{2009ASPC..415...87L}) is a 3D non-LTE radiative transfer code that treats the radiative transfer computations using the \textit{accelerated lambda iteration} (ALI) method developed by \citet{1992A&A...262..209R}. We computed the radiation field in full 3D where, for each grid point, angle integration for rays in 24 directions (3 angles in each octant) were carried out using the A4 ray set provided by \citet{1963...1C}.  In order to increase stability of the runs, the patch by \citet{2019A&A...631A..33B} was used. The electron densities were calculated assuming the local thermodynamic equilibrium (LTE)  and using the opacity package included in the Multi3D code. \\
\newline
Every forth snapshot is used as input for the computations. To reduce the computational time, we crop the model in two ways. The vertical span of the input model is limited to $360$ point which includes the range of temperatures at which the spectral lines that we generate are formed.  In the horizontal direction, we use every second grid point to make computations faster. Tests show that this does not affect the appearance of synthetic features. % The code is able to handle effects of partial frequency redistribution using \textit{multilevel accelerated lambda iteration} (M-ALI) \citet{2001ApJ...557..389U},  but complete frequency redistribution was used in this simulation.

\subsection{Model atoms}
We used two model atoms for our computations; hydrogen and silicon.\\
\newline
H$\alpha$ was calculated using a 3-level plus continuum H model atom. The model was used by \citet{2012ApJ...749..136L}, and modified by \citet{2019A&A...631A..33B} who removed the $n=4$ and $n=5$ levels to increase numerical stability and ensure convergence of the \textit{multilevel accelerated lambda iterations} (M-ALI) in Multi3D. Processes included in the model are excitation from electron collisions with positive ions, and ionisation by electron collisions.\\
\newline
The Si~IV model atom includes 3 ionization levels (Si~III, Si~IV and Si~V) in order to synthesize the $1394$~\AA\ and $1403$~\AA\ lines which originate in the $2p^6 3s {}^2 S_{1/2}$-$2p^6 3p {}^2P_{3/2}$ and $2p^63s {}^2S_{1/2}$-$2p^6 3p {}^2P_{1/2}$ transitions, respectively. All oscillator strengths $f_{ij}$ and energy levels were taken from the NIST database (\citet{NIST_ASD}). The abundance of Si with respect to hydrogen was set constant as $A_{Si}=7.51$, as per \citet{2009ARA&A..47..481A}. We used the photospheric abundance as it has been shown that low-FIP elements can have photospheric composition in the upper atmosphere during impulsive heating events (\citet{2016ApJ...824...56W}). However it could be beneficial to investigate coronal abundances in similar computations in the future to see the impact of different assumptions as the line intensity of the $1403$~\AA\ line has been shown to become larger when coronal abundances were used (see e.g.\ \citet{2019ApJ...871...23K}, \citet{2003SSRv..107..665F}). In the Si~IV model we include excitation from electron collisions with positive ions, collisional ionisation, collision auto-ionisation, charge transfer with neutral hydrogen atoms, electron impact ionisation of complex ions and radiative recombination. In the computations, the dielectronic recombination rate is suppressed (using a subroutine in Multi3D which follows the work of \citet{1974MNRAS.169..663S}) as it is an effect that takes place for high electron densities in which ranges our results lie. Density sensitive dielectronic recombination can lead to an elevated Si~IV fraction. The effect of including dielectronic auto-ionisation and secondary auto-ionisation corrections is clearly visible for Si~I to V (see Figure 14 in Appendix B of \citet{2019ApJ...871...23K}). The coefficients for all included processes that were used were taken from the Si~IV model atom used in \citet{2019A&A...627A.101V}.\\
\newline
The Multi3D computations are done assuming statistical equilibrium and \textit{complete frequency redistribution} (CRD, \citet{2012ApJ...749..136L}). In Fig.~\ref{fig:intro} we present the full FOV in a) H$\alpha$ blue wing at 53~km~s$^{-1}$, b) H$\alpha$ line core intensity, c) intensity of Si~IV $1394$~\AA\ line summed over all frequencies $1393.755\pm0.79$~\AA\. The white box indicates the ROI. 
% no stark broadening
% separate radiative transfer section?
\begin{figure*}
   \centering
   \includegraphics[width=180mm]{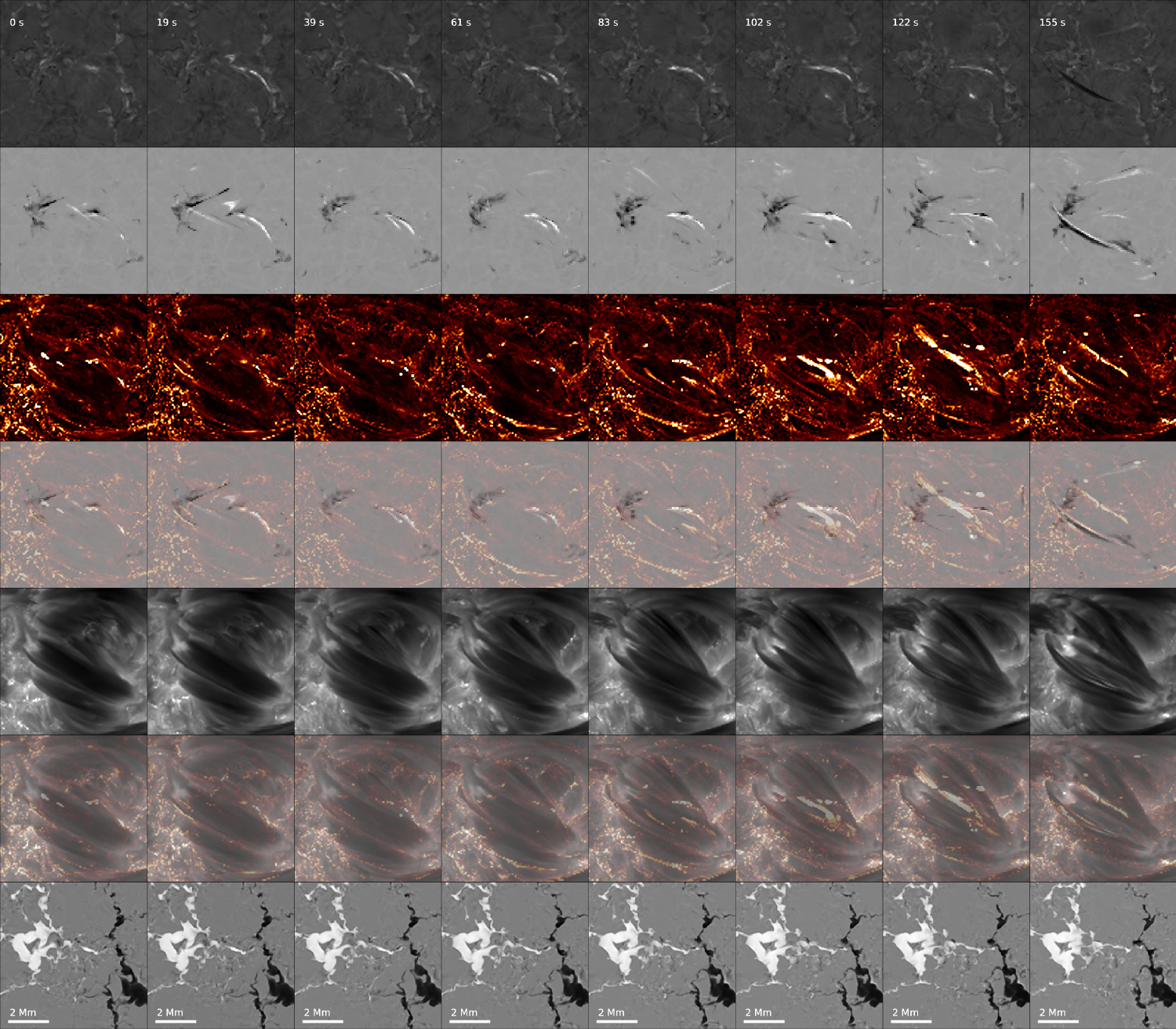}
   \caption{Time evolution of the synthetic loop at $\mu=1$. From the top to the bottom: normalized H$\alpha$ blue wing intensity at $v_D = 53$~km~s$^{-1}$, H$\alpha$ Dopplergram at velocities $\pm53$~km~s$^{-1}$  (black/white represent red-/blue wings), Si~IV~$1393$~\AA\ summed over frequencies $\pm0.79$~\AA\ around line center, composite image of the H$\alpha$ Dopplergram (black/white) and frequency summed Si~IV (orange), H$\alpha$ line core, composite image of H$\alpha$ line core and frequency summed Si~IV and vertical magnetic field (black/white represents downward/upward magnetic field).}          
    \label{fig:1.1}%
\end{figure*}
\begin{figure}
   \centering
   \includegraphics[width=0.8\linewidth,trim= 1.6cm 0cm 1.6cm 0cm,clip=true]{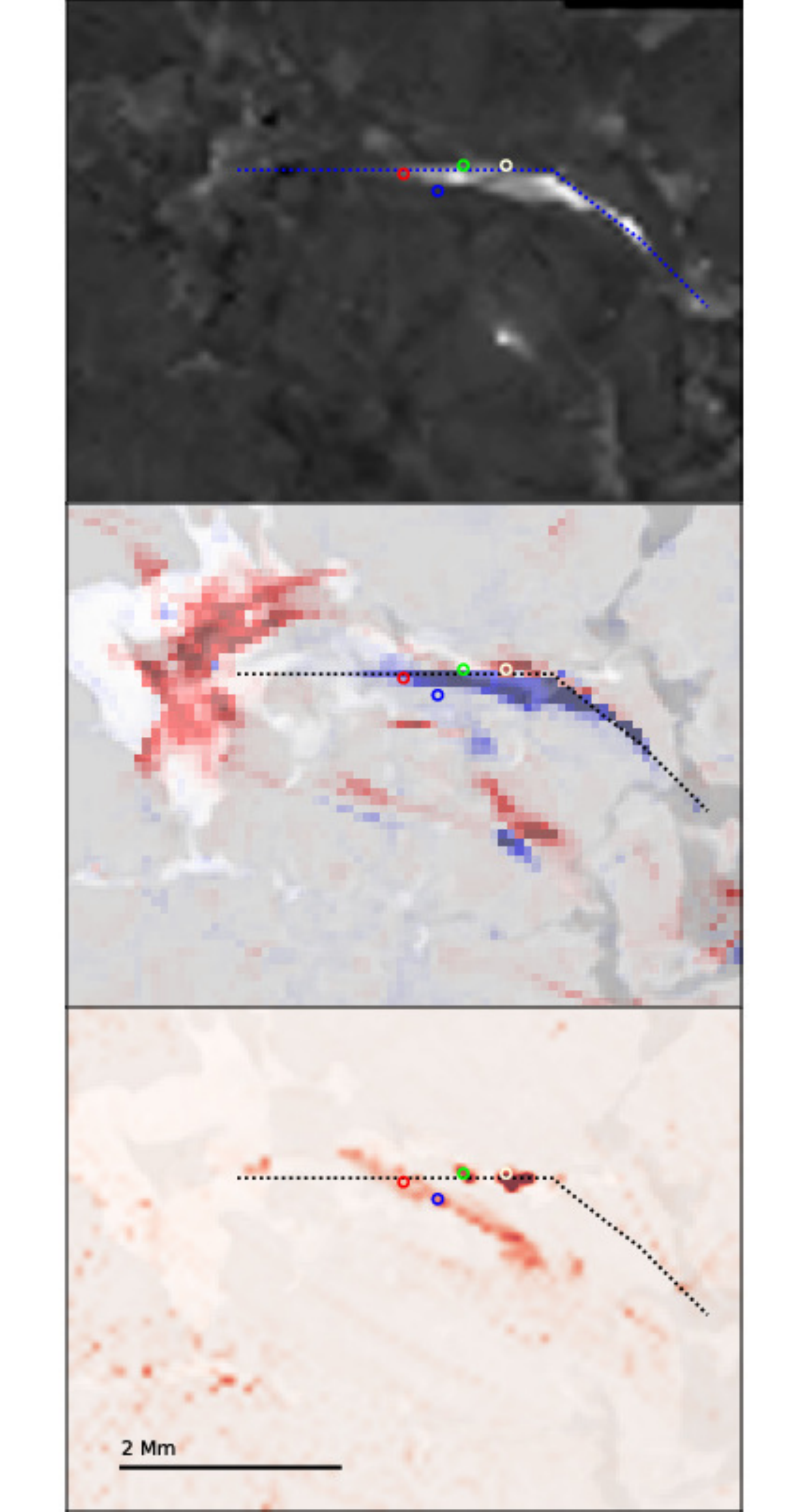}
   \caption{H$\alpha$ blue wing intensity (top row), H$\alpha$ Dopplergram (middle row) and Si~IV frequency-integrated intensity (bottom row) in the snapshot at $t=102$~s. Three pixels of interest are highlighted using red, green and white circles. Dotted line marks the position of the vertical cut shown in Fig.~\ref{cut}. }
    \label{fig:3}%
\end{figure} 
\begin{figure}
   \centering
   \includegraphics[width=\linewidth,trim= 0.65cm 0.45cm 0.6cm 0.4cm,clip=true]{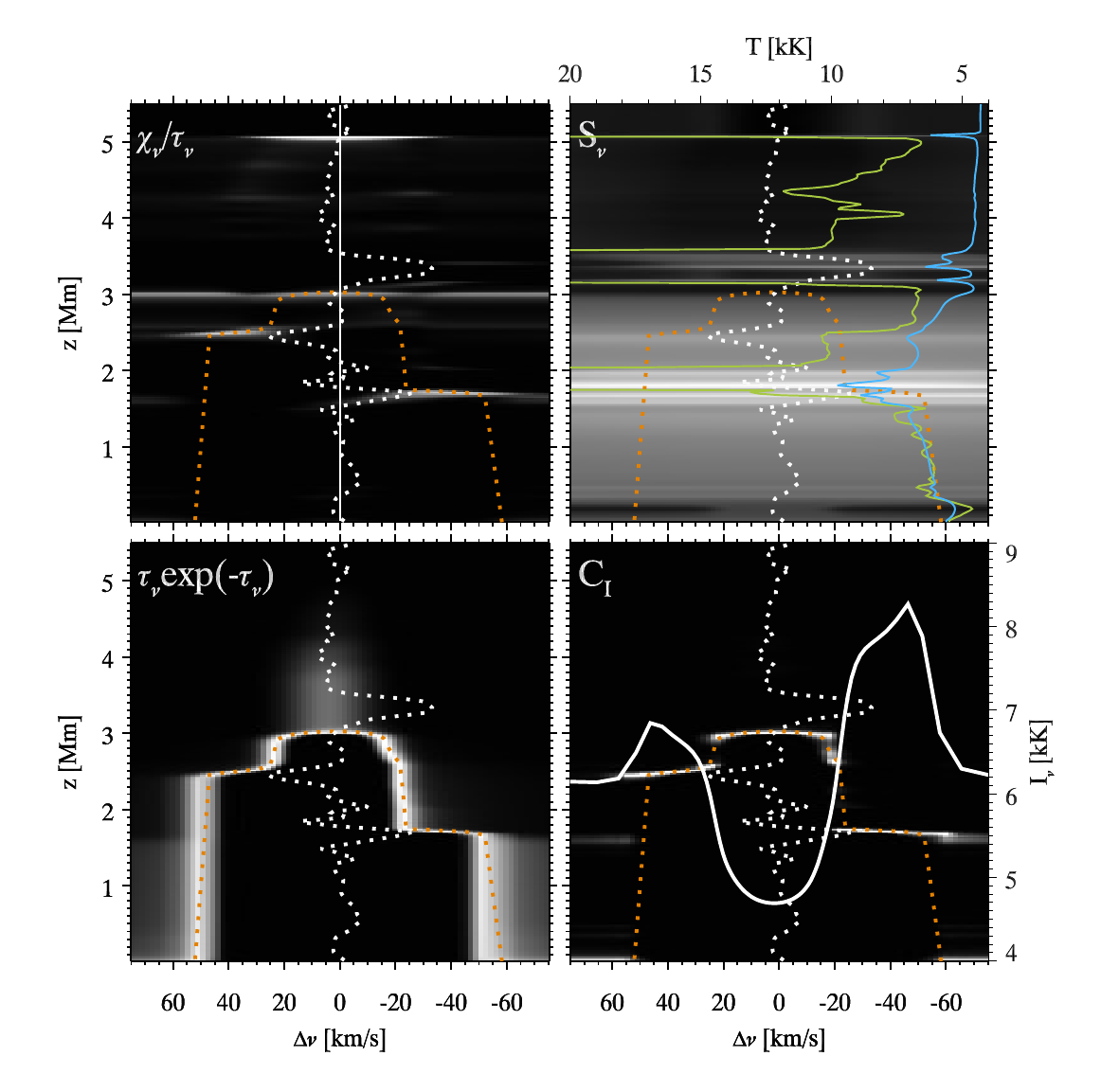}
   \caption{H$\alpha$ $\mu=1$ four-panel diagram at $t=102$~s in the pixel indicated with a white circle in Fig.~\ref{fig:3}. The plotted quantities, as functions of frequency and altitude, are indicated in the upper left corner of each subplot and plotted as a grayscale. $z=0$~Mm marks the photosphere. In all panels the vertical velocity as function of altitude is shown (white dotted line), where upwards velocity is positive and downwards is negative. In the upper left panel $v_z = 0$ is indicated by a vertical solid white line. Additionally, the altitude where $\tau=1$ is also plotted in all subplots (orange dotted line). The different quantities are as follows: 
   \textit{Upper left:} opacity $\chi_\nu$ over optical depth $\tau_\nu$,    
   \textit{upper right:} total source function $S_\nu$ (Eq. \ref{eq:Stot}), temperature (green line) and source function (blue line),
   \textit{lower left:} attenuation factor $\tau_\nu exp(-\tau_\nu)$,
   \textit{lower right:} contribution function $C_I$ (Eq. \ref{eq:CI}), and emergent intensity (solid white line).
   }
        \label{fig:4.1}%
\end{figure}
\begin{figure}
   \centering
   \includegraphics[width=\linewidth,trim= 0.65cm 0.45cm 0.6cm 0.4cm,clip=true]{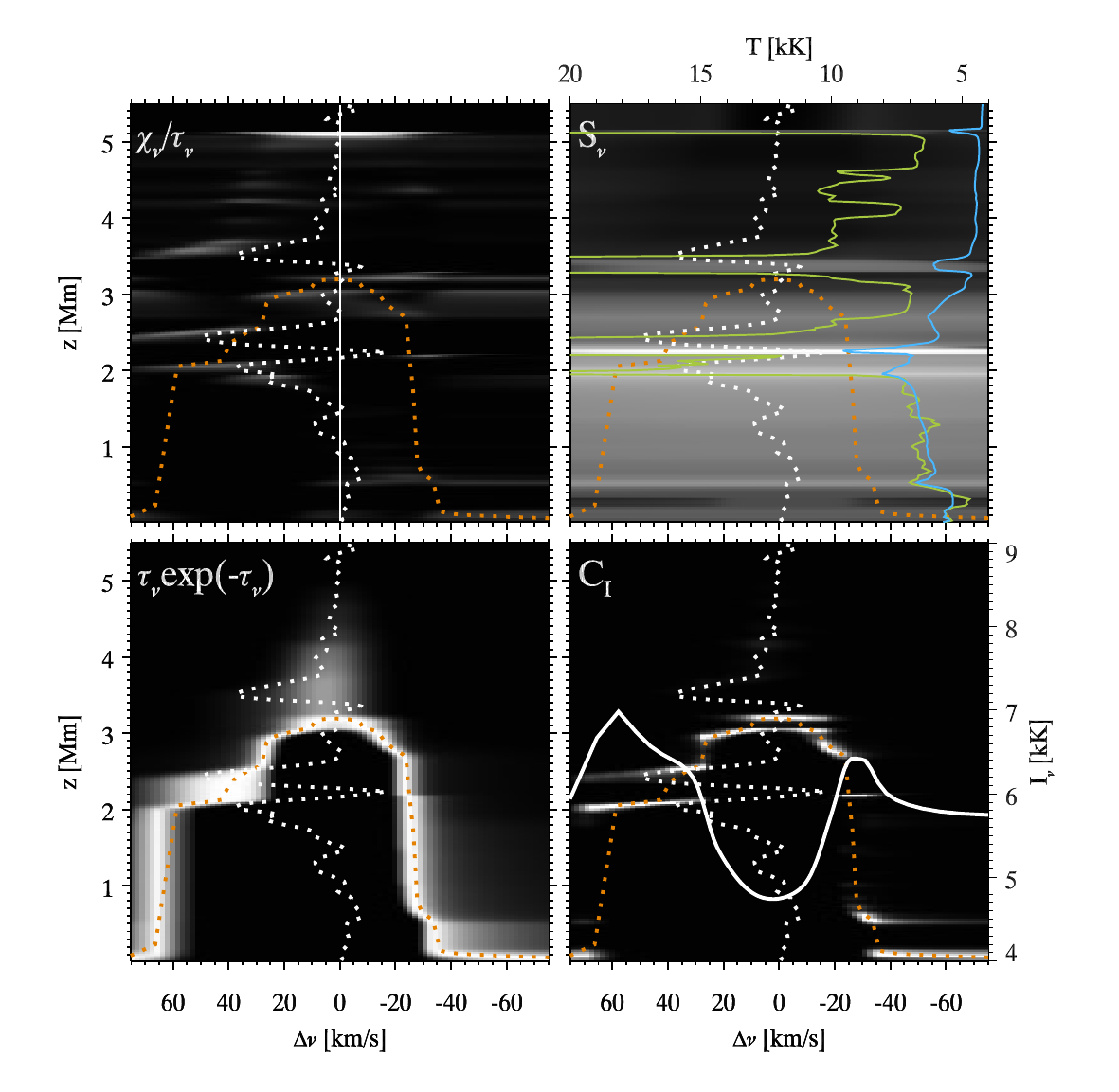}
   \caption{H$\alpha$ $\mu=1$ four-panel diagram at $t=102$~s in the pixel indicated by the green circle in Fig.~\ref{fig:3}. The format of the diagram is the same as in Fig.~\ref{fig:4.1}.
   }
        \label{fig:4.2}%
\end{figure} 
\begin{figure}
   \centering
   \includegraphics[width=\linewidth,trim= 0.65cm 0.45cm 0.6cm 0.4cm,clip=true]{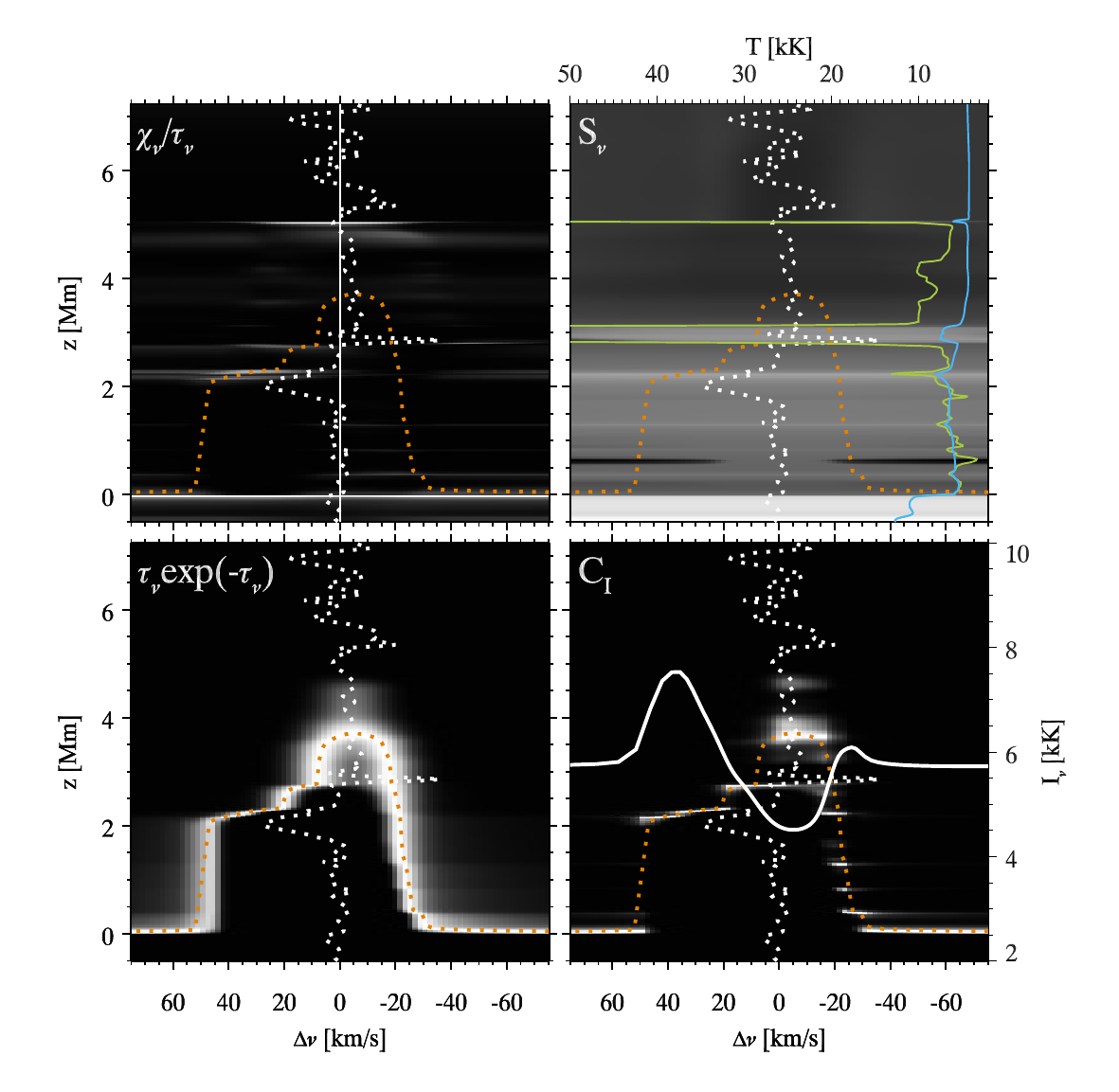}
   \caption{H$\alpha$ $\mu=1$ four-panel diagram at $t=102$~s in the pixel indicated by the red circle in Fig.~\ref{fig:3}. The format of the diagram is the same as in Fig.~\ref{fig:4.1}.
   }
        \label{fig:4.3}
\end{figure} 
\section{Results}
\label{s:results}
Further on, we focus on ROI and show synthetic observables at two viewing angles $\mu=1$ and $\mu=0.66$. In the final subsection, we study the field evolution that leads to features formation.

\subsection{Synthetic structure at $\mu=1$}
\subsubsection{Evolution of synthetic structure at $\mu=1$}
\begin{figure*}[h]
   \centering
   \includegraphics[width=180mm]{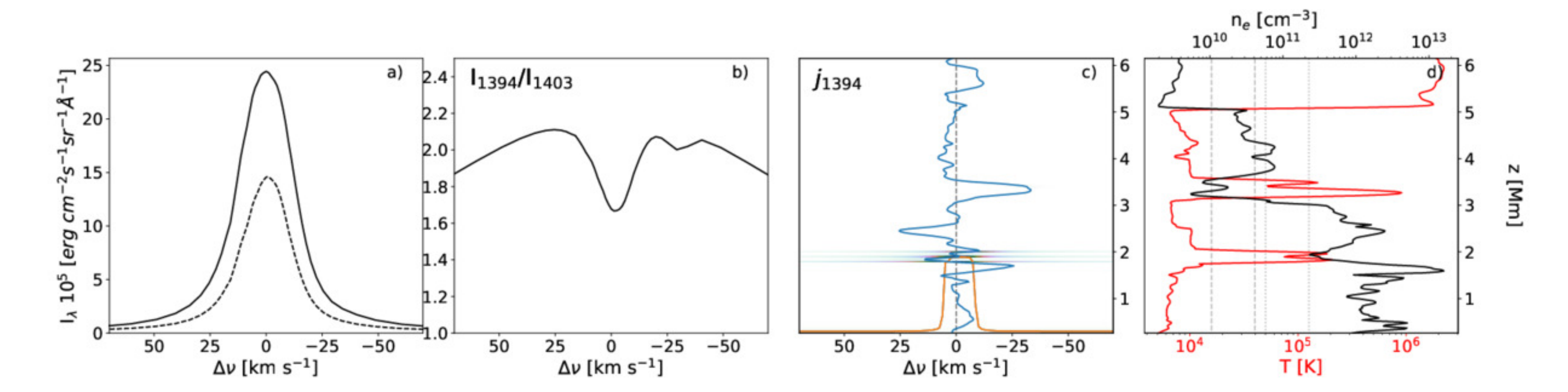}
   \caption{a) Line profiles of Si~IV 1394 $\AA$ (solid black) and Si~IV~$1403$~\AA\ (dashed black) in the pixel marked by a white circle in figure \ref{fig:3}. b) Intensity ratio of Si~IV~$1394$~\AA\ and Si~IV~$1403$~\AA\ lines. c) Emissivity $j$ of the $1394$~\AA\ line (eq. \ref{eq:jnu}, $j^{0.6}$) and vertical velocity $v_{z}$ (blue). The vertical dashed line indicates where $\Delta v=0$~km~s$^{-1}$. The orange line marks the height at which $\tau=1$ is reached. d) Gas temperature $T$ (red) and electron density $n_e$ (black) of Si~IV. The dashed vertical lines represents the LTE Si~IV formation temperature interval and the dotted vertical lines represents the coronal equilibrium formation temperature interval. Positive $\Delta \nu$ indicates an upwards velocity.}
              \label{fig:6.1}%
\end{figure*}

\begin{figure*}[h]
   \centering
   \includegraphics[width=180mm]{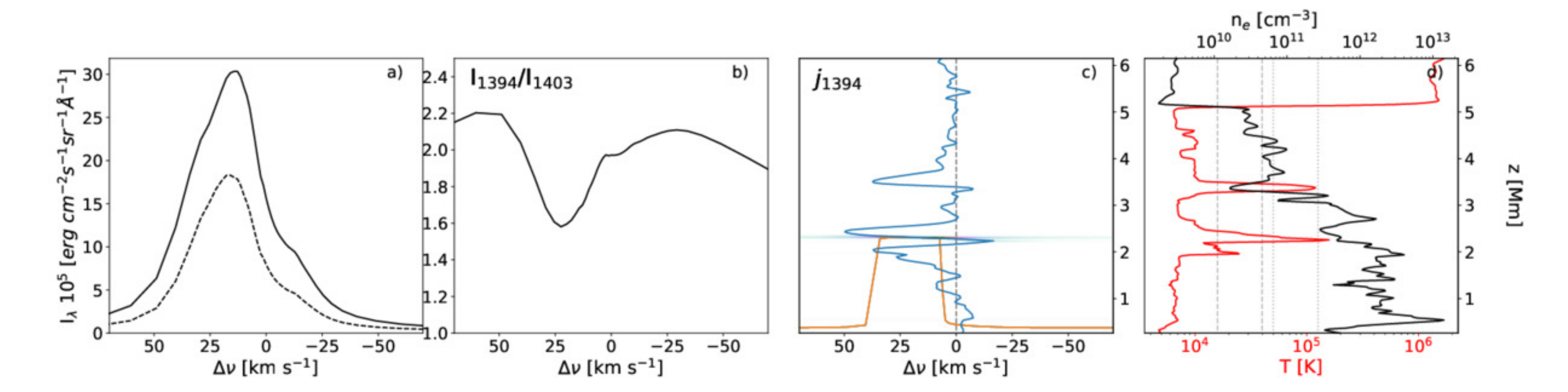}
   \caption{The same layout as in Figure \ref{fig:6.1}, but for the pixel marked by a green circle in Figure \ref{fig:3}.}
              \label{fig:6.2}%
\end{figure*}

\begin{figure*}[h]
   \centering
   \includegraphics[width=180mm]{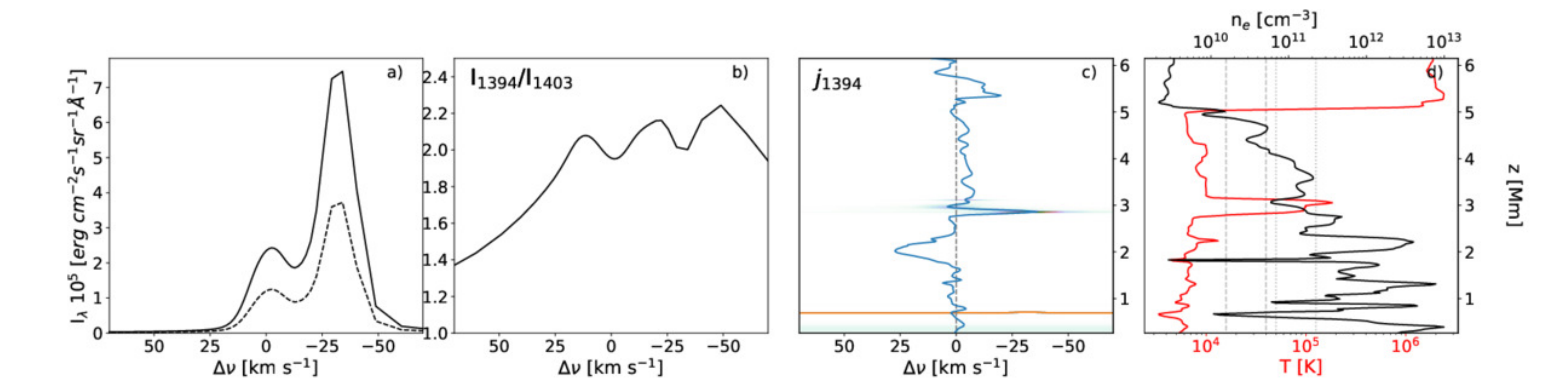}
   \caption{The same layout as in Figure \ref{fig:6.1}, but for the pixel marked by a red circle in Figure \ref{fig:3}.}
              \label{fig:6.3}%
\end{figure*}

\begin{figure*}[h]
   \centering
   \includegraphics[width=180mm]{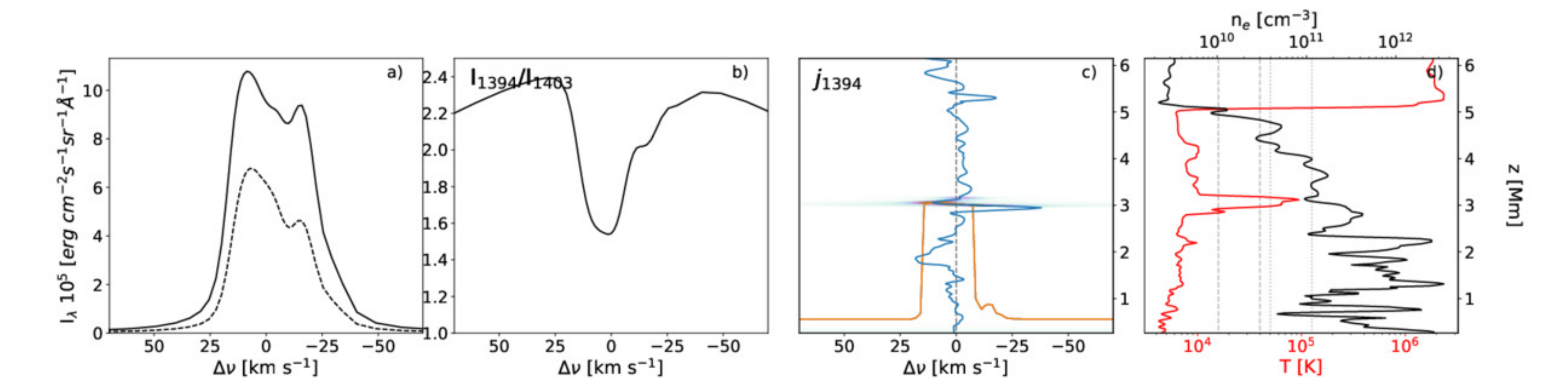}
   \caption{The same layout as in Figure \ref{fig:6.1}, but for the pixel marked by a blue circle in Figure \ref{fig:3}.}
              \label{fig:6.4}%
\end{figure*}

In Figure \ref{fig:1.1}, the time evolution of the resulting synthetic structure is shown at $\mu=1$. The upper row of the figure shows the blue wing of the H$\alpha$ line. In the second row we have constructed a Dopplergram in order to showcase the strong Doppler shifts in the H$\alpha$ wings. The Dopplergrams were created by subtracting the red wing from the blue, and dividing it by the sum of the two such as  
\begin{equation}
    I_D = \frac{I_B - I_R}{I_B + I_R}
    \label{eq:ID}
\end{equation}
The third row from the top displays the Si~IV emission, summed over frequency. The fourth row is a composite of the Dopplergram and the Si~IV emission. The fifth row shows the H$\alpha$ line core and the sixth is a composite of the summed Si~IV intensity and the H$\alpha$ line core. The bottom row displays the vertical magnetic field $B_z$ at a photospheric level, where black/white indicates negative/positive direction of the field.\\
~\newline
The figure shows time period of $155$~s during which we see appearance and disappearance of one complex feature visible in H$\alpha$ Dopplergams. Throughout the entire evolution, this loop-like structure has a predominantly positive $I_D$. In the first two snapshots, a jet-like feature is visible emanating from the positive-polarity region. No structure can be seen in the H$\alpha$ blue wing at $t=0$ s. Over $20$~s, a loop-like structure appears in the blue wing which increases in brightness and grows to about $2$~Mm in length. At $39$~s, the loop-like structure forms a fork with a smaller branch staying visible until $t=83$~s. Up until this point, the loop has mainly positive $I_D$, but at $t=83$~s the $I_D$ at the top of the loop becomes negative. The entire structure grows in complexity, i.e.\ multiple strands of red- and blueshifts are created and disappear at a fast rate up until $t=122$ s. During the same time, the width of the loop-like structure grows until $102$~s and then shrinks. At the very end of the run, at $t=155$ s, the loop-like structure is almost completely gone. Instead, slightly shifted and in a different direction, a long loop with predominantly negative $I_D$ appears.\\   
~\newline
%In the line core of H$\alpha$ we see a structure similar to a small arcade. The strands of the structure are, in the beginning of the run, tight knit. At the end of the run, the structure has taken on a more loose appearance consisting of thin filaments. 
The Si~IV evolution is different from the one in H$\alpha$. Increased emission is visible in almost the whole ROI. The bright points are distributed in the regions with a strong vertical magnetic field. In parts where the magnetic field is more horizontal, the Si~IV emission is concentrated in stripes that largely follow the fibrils visible in the H$\alpha$ line core. Here and there, very bright spots form and disappear. Some of them overlap with the loop-like shape visible in the H$\alpha$ Dopplergrams. At $t=102$~s some overlapping bright points remain, but a straight shape forms in a different spatial plane than the loop. For $20$~s, this feature grows by approximately $1$~Mm in length diagonally over the H$\alpha$ loop-like structure. In the last snapshot, the same feature is shrunk in thickness, length and brightness. A new, smaller elongated feature appears parallel to the loop-like structure visible in H$\alpha$. The Si~IV-H$\alpha$ line core composite image shows a behaviour where the Si~IV emission seemingly follows the hydrogen line core structure between $83-155$~s. \\
~\newline
In Fig.~\ref{fig:3}, we chose to highlight the snapshot at $t=102$~s. In the top row the synthetic H$\alpha$ blue wing intensity is presented. The middle row shows H$\alpha$ Dopplergram created in the same manner as in Fig.~\ref{fig:1.1}, but with the difference that $I_D>0$ and $I_D<0$ are here represented by blue and red colours, respectively. The Dopplergram is over-plotted over the photospheric vertical magnetic field. In the bottom row, the Si~IV frequency-summed intensity is plotted over the same field map. The white, green and red circles that appear in all subplots marks three pixels of interest that we analyse using four panel diagrams in Figs.~\ref{fig:4.1}-\ref{fig:4.3} as well as using the Si~IV emissivity $j$ in Figs.~\ref{fig:6.1}-\ref{fig:6.4}, including the pixel marked with the blue circle. Three of four chosen pixels are in regions where the H$\alpha$ loop-like structure and Si~IV emission overlap. We see in the middle row of Fig.~\ref{fig:3}, a complex structure in red to the left of the loop at the positive pole of the bipolar region, as well as the feature below the loops where Doppler shifts of both sign are present. The H$\alpha$ loop like-structure is stretched between the two magnetic poles with a slight curvature, making it bulge out from the region directly between the two. The footpoints do not reach the poles in this Dopplergram at $\pm 53$~km~s$^{-1}$. However, Dopplershifts at $\pm 39$~km~s$^{-1}$ does indeed show a loop-like structure with footpoints overlapping with the magnetic poles. In the case of Si~IV, the elongated structure in emission is spanning between the poles in an approximately straight line. The straight structure crosses the H$\alpha$ Dopplergram loop-like structure but is not parallel to it, it is located in a different spatial plane.  
\subsubsection{H$\alpha$ line formation}
In Figures \ref{fig:4.1}, \ref{fig:4.2} and \ref{fig:4.3} we analyse the three pixels indicated by circles in Fig.~\ref{fig:3} in more detail using four-panel schematic H$\alpha$ line formation diagrams \citep[as introduced by][]{1997ApJ...481..500C}. Each figure is divided into four subplots. In all of the subplots the dotted orange line represents $\tau=1$, and the dotted white line the line of sight velocity $v_z$. Positive $\Delta v$ marks an upwards flow of the plasma. In the upper left subplot we show how the opacity divided by optical depth, $\chi_\nu / \tau_\nu$, is changing with Dopplershift $\Delta v$ and altitude $z$. Regions with high opacity and low optical depth are visible as bright areas in the plot, they indicate that the density of H$\alpha$ emitting particles at that specific height and Dopplershift is higher than its surroundings. The straight solid line indicates where $\Delta v=0$ km~s$^{-1}$. The upper right subplot displays the source function $S_\nu^{tot}$ as a grayscale. $S_\nu^{tot}$ is the total source function which combines the line source function $S_\nu^l$ and the continuum source function $S_\nu^c$;
\begin{equation}
    S_\nu^{tot} = \frac{j_\nu^l + j_\nu^c}{\alpha_\nu^l + \alpha_\nu^c}
    \label{eq:Stot}
\end{equation}
where $j_\nu^l$ and $j_\nu^c$ are the line/continuum emission coefficients, and $\alpha_\nu^l$ and $\alpha_\nu^c$ are the line/continuum extinction coefficients \citep[see ][]{2003rtsa.book.....R}. Additionally, line source function $S_\nu^l$  expressed as excitation temperature and the
gas temperature are plotted as a blue and green line, respectively. The lower left subplot shows the attenuation factor $\tau_\nu e^{-\tau_\nu}$. 
Finally, the lower right subplot shows the contribution function $C_I = dI_\nu /dz$, i.e.\ the contribution to the H$\alpha$ emergent intensity. $C_I$ is here calculated as 
\begin{equation}
    C_I = \frac{\chi_\nu}{\tau_\nu} \tau_\nu e^{-\tau_\nu} S_\nu^{tot}.
    \label{eq:CI}
\end{equation}
The intensity profile is over-plotted as a white solid line. \\
\newline
Figure~\ref{fig:4.1} is created from a strongly Dopplershifted pixel in the region marked by a white circle in Fig.~\ref{fig:3}. In the upper left panel we see that $\chi_\nu / \tau_\nu$ is higher in regions that experience strong gradients in up- and downflows where $\tau = 1$. In the regions around $1.7$ and $2.5$~Mm there is an increased $\chi_\nu / \tau_\nu$ that is broadened in both blue and red line wings as a consequence of the strong velocity gradients. At about $3$~Mm there is emission that is almost constant for all considered frequencies. This area is exposed to a downflow from above. The red wing is formed at an altitude of $1.7$~Mm where the plasma has a downward velocity. The blue wing is formed in a region which is located higher up at an altitude of about $2.4$~Mm in material that is travelling upwards. The line core is formed at $3$~Mm, slightly below a strong downflow and a hot cell in the upper chromosphere. In the upper right panel the line source function is constant for all frequencies as we assume CRD. We see that two hot pockets are formed in the atmosphere at altitudes $1.75-2$~Mm and $3.1-3.6$~Mm, and that the transition region is located slightly below $5.1$~Mm. $S^l_{\nu}$ follows the temperature fairly well from the photosphere up to $1.75$~Mm, where it then decouples from the Planck function in the lower hot chromospheric pocket. Some spikes in the source function appear in this region as a result of the high density of hot plasma in the area. Between the two hot pockets $S^l_{\nu}$ globally declines in strength, with exception for a peak at $2.4$~Mm, coinciding with an upflow of the local material. In the upper hot pocket, the source function experiences three additional peaks. The lower left plot displays the area where $\tau$ is close to $1$. The strongest values of $\tau_\nu e^{-\tau_\nu}$ is obviously centered around the orange dotted $\tau=1$ curve, but there is a broad bright area between $3-4.2$~Mm centered around zero Dopplershift. Additionally, there is a symmetrical and strong broadening in the wing forming region and below. From the contribution function plot at the lower right we can see that the line core is formed at $3$~Mm, the blue wing at $2.5$~Mm and the red wing around $1.7$~Mm. Both of the wings are in emission and the core in absorption. There is an asymmetry in the wings where the red part has a higher intensity than the blue counterpart. The line core is approximately symmetric. The wing asymmetry originates from the peak in the line source function at $1.7$~Mm which is higher than the peak at the altitude in which the blue wing is formed. Both of the wings display a "knee" which are formed in regions that experience a strong velocity gradient. \\
\newline
The four-panel diagram created in the pixel indicated by a green circle in Fig.~\ref{fig:3} is presented in Fig.~\ref{fig:4.2}. This pixel experiences a much more complex opacity field than in the case of Fig.~\ref{fig:4.1}. Here, we have numerous bright spots in the upper left panel at a variety of altitudes and frequencies in contrast to the previous case where broad profiles are centered in a few regions with strong velocity gradients. A similar feature between the two four-panel diagrams is that both have emission at the upper chromosphere/lower transition region. 
There are two hot pockets along the column which are overlapping with the peaks in upflowing vertical velocity and peaks in the line source function. $S^l_{\nu}$ follows the temperature curve in the lower chromosphere up until it reaches the lower hot pocket at an altitude of $1.9$~Mm where NLTE decoupling occurs. We can, just like in Fig.~\ref{fig:4.1}, see broadening in $\tau_\nu e^{-\tau_\nu}$ which in this case mainly occurs between $2.1-2.5$~Mm where the plasma experiences a strong upflow. Line wing and core intensities are all experiencing contribution from more than one altitude. In the line core, the lower line formation region at an altitude just above $3$~Mm is blueshifted, causing the line core to become slightly asymmetric as it is absorbing some of the photons emitted by the blue wing which is formed below the core. Similar to the profile in Fig.~\ref{fig:4.1}, both of the wings in the line profile are in emission. This pixel also displays a knee, but here only in its blue wing. \\
\newline
Finally, in Fig.~\ref{fig:4.3} we show the four-panel diagram of the pixel indicated by a red circle in Fig.~\ref{fig:3}. The pixel is located where the loop-like shape in the H$\alpha$ blue wing is overlapping with the larger Si~IV structure that stretches across the area. In contrary to the two other pixels, there is not much broadening of $\tau_\nu e^{-\tau_\nu}$ due to the fact that the velocity does not have such a complex profile when compared to Figs.~\ref{fig:4.1} and \ref{fig:4.2}. There is a peak in upwards velocity around $2$~Mm and a peak in downflow slightly below $3$~Mm. In the lower right panel we see that the line profile of H$\alpha$ is strong in its blue wing and that the line core is slightly asymmetric. When investigating the contribution function $C_I$ it is obvious that the strong blue wing has its origin at $\approx 2.2-2.5$~Mm in an area with a peak in temperature, just above a upflowing region. The red wing is hardly in emission, and it finds contribution from a range of altitudes. The line core is slightly redshifted, causing it to absorb some of the red wing photons that are emitted at a lower altitude. There is contribution to the line core between $\approx2.5-5$~Mm, with its largest contribution coming from a range between $3-4.2$~Mm. The line profile does not display a knee in any of its wings like it does in the two prior four-panel diagrams, which is due to the lack of harsh velocity gradients.\\
~\newline
\subsubsection{Si~IV line formation}
The Si~IV lines are considered to be optically thin, in which case the approximate formation height of the line is where the emissivity is at its maximum. The emissivity can be calculated as
\begin{equation}
    j_\nu = S_\nu^{tot} \chi_\nu.
    \label{eq:jnu}
\end{equation}
Figures \ref{fig:6.1}-\ref{fig:6.3} are composed of four subplots with quantities from the pixels marked by circles in Fig.~\ref{fig:3}. In subplot a) the line profiles of Si~IV~1394~$\AA$ (solid black line) and Si~IV~1403~$\AA$ (dashed black line) are plotted together and subplots b) shows their ratio. Subplot c) shows the emissivity $j_\nu$ of the Si~IV~1394~$\AA$ and the vertical velocity $v_z$ (blue solid line) as they vary with Dopplershift and altitude. The orange solid line marks the height at which $\tau=1$. The grey dashed line indicates where $\Delta v=0$ km s$^{-1}$. The emissivity of 1403~$\AA$ line is very similar in all cases, so we omit plotting it. In subplot d) the gas temperature $T$ (red) and electron density $n_e$ (black) are plotted as functions of height. The grey dashed lines indicates the interval in which LTE Si~IV formation occurs ($10^{4.2}$-$10^{4.6}$ K), and the dotted vertical lines represents the interval of coronal equilibrium Si~IV formation ($10^{4.7}$-$10^{5.1}$ K) (see e.g.\ Figure~4 in \cite{2016A&A...590A.124R}). \\
~\newline
In Figure~\ref{fig:6.1}, the two line profiles are approximately symmetrical around their respective line cores in subplot a). We see a high emissivity at $z=2$ Mm and slightly below, appearing in one thicker band at $1.9$~Mm and two thinner bands at $2$ and $1.8$~Mm. These three bands in the emissivity coincide with locations where temperature is around $10^{4.8}$~K. The symmetrical shape of the line profiles is due to the presence of symmetric moderate flows in both directions, both up- and downflows of $10-20$~km~s$^{-1}$ in the area. The emissivities are very similar in both $1394\AA$ and $1403\AA$ and the lines are formed in the same regions. This is true for Figs.~\ref{fig:6.2} and \ref{fig:6.3} as well.\\ % \textcolor{red}{The increase in width is thus a product of broadening due to Dopplershift. Pure Doppler broadening is valid in regions with high temperatures and low mass densities, which is the case at this altitude.} 
\newline
Figure~\ref{fig:6.2} shows a broad peak in $j_\nu$ at $2.3$~Mm that is overlapping with a high temperature and the electron density of $10^{12}$~cm$^{-3}$. In the region where the emissivity is peaking the velocity gradient is steep with the maximum upflow velocity of $\approx50$~km~s$^{-1}$. A smaller peak $j_\nu$ at $\approx2.2$~Mm overlaps with a small downflow of $-10$~km~s$^{-1}$. As a result, the line profiles are asymmetric, largely blueshifted, with a slightly extended red wing. \\
\newline
Figs.~\ref{fig:6.3} and ~\ref{fig:6.4} show the properties of the pixels located on the larger structure in Si~IV. In Figs.~\ref{fig:6.3}, the line profiles are highly asymmetric with two peaks. The smaller peak is located around $\Delta v=0$~km~s$^{-1}$ and the larger is redshifted to a velocity of $-34$~km s$^{-1}$. In the emissivity diagram we see emission peak centred around $3$~Mm which is coinciding with a peak in temperature and the electron density of less than $10^{11}$~cm$^{-3}$. Most of the emission occurs in a downflowing region, giving rise to the almost completely redshifted line profiles. The smaller peak in the line profiles comes from the region slightly above, where the vertical plasma velocity is close to $\Delta v=0$~km~s$^{-1}$. The position of this hot pocket is better visible in Figs.~\ref{cut} and ~\ref{fig:fh_vapor}. Figure~\ref{fig:6.4} corresponds to the pixel that is further down the straight Si~IV feature visible in Fig.~\ref{fig:3}. Temperature peak is at the same height as in Figs.~\ref{fig:6.3}, but now the corresponding electron density is $60$~$\%$ higher. The emission peak is located in the region at temperature of $\approx 10^{4.7}$~K where the velocity jumps from downflow of $\approx50$~km~s$^{-1}$ to the $\approx20$~km~s$^{-1}$ upflow. This gives a relatively symmetric line profiles. The secondary emission peak in line profiles comes from variations in the electron density.\\   
\newline
To check whether the optically thin assumption holds, we overplot the height at which optical depth unity is reached and calculate the ratio of the intensities of the two lines. These are shown in subplots b) and c) of Figs.~\ref{fig:6.1}-\ref{fig:6.4}. In the optically thin regime, the ratio of the intensity profiles of Si~IV is equal to the rate of their oscillator strengths which is exactly $f_{1394}/f_{1403}=2$. Also, the optical depth unity level in that case should be far from the locations of strong emission. In all chosen pixels, the line ratio shows wavelength dependence with values going larger than 2 in line wings. However, when we look at the $\tau=1$ level, the wavelength range where its location is close to the emission peak is very narrow and corresponds to the location where the line ratio is $1.6$. Only in case shown in Figs.~\ref{fig:6.3}, the Si~IV lines seems to be formed in optically thin regime.
If we calculate the wavelength integrated ratio of the two lines
\begin{equation}
    R = \frac{\int I_{1394}(\lambda) d\lambda}{\int I_{1403}(\lambda) d\lambda},
\end{equation}
we get values between 1.6 and 1.8, so $R<2$ in all four cases.
\begin{figure*}
   \centering
   \includegraphics[width=180mm]{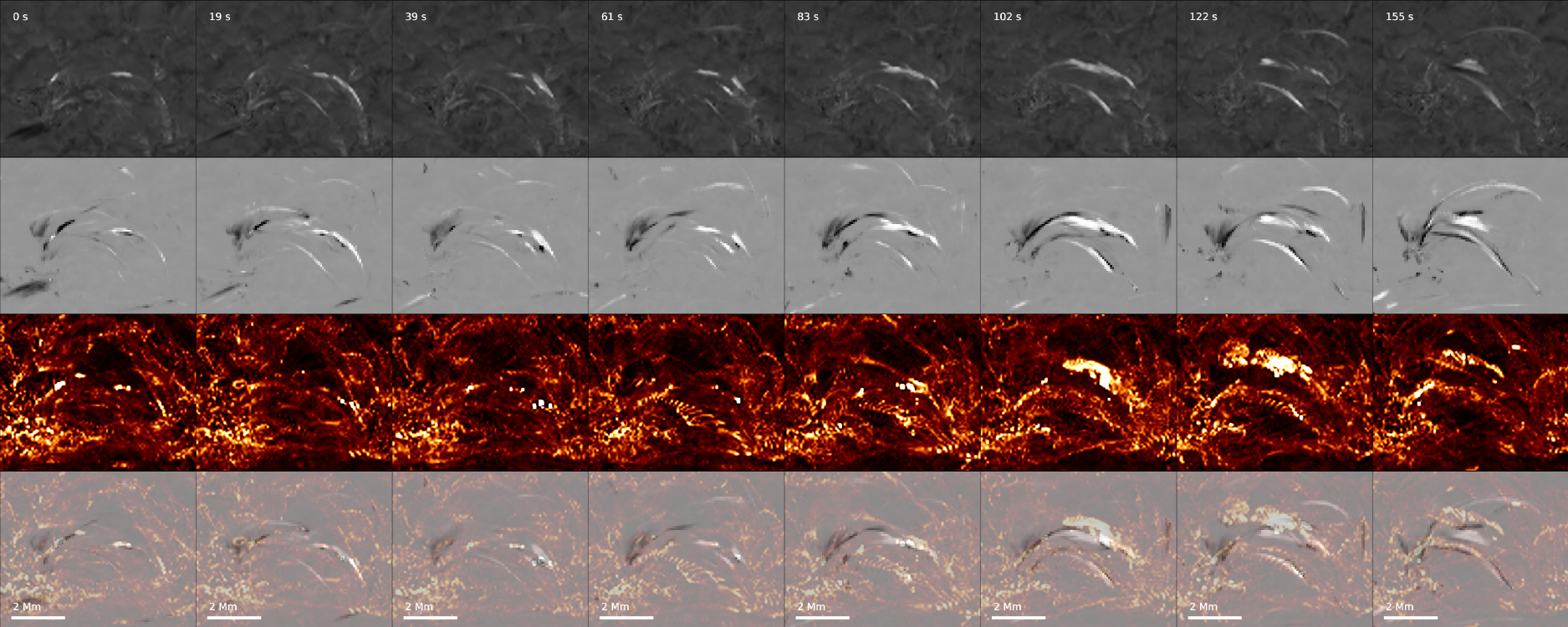}
   \caption{Time evolution of the synthetic loop at $\mu=0.66$. The rows show the same quantities as in Fig.~\ref{fig:1.1}, minus the vertical magnetic field and the H$\alpha$ line center.}
              
    \label{fig:1.2}%
\end{figure*} 
\subsection{Synthetic structure at $\mu=0.66$}
Figure \ref{fig:1.2} shows the temporal evolution of ROI at $\mu=0.66$. The figure outline is the same as in Fig.~\ref{fig:1.1}, without the rows that show the vertical magnetic field and the H$\alpha$ line center. H$\alpha$ Dopplergrams show not one, but two loop-like structures in this slighted view. The upper loop is the same one that is visible at $\mu=1$, but the structure has a more defined loop shape consisting of thin strands relative to the case of $\mu=1$. The half-length of the loop-like structure is roughly $2.4$~Mm.
At $t=0$~s, there are some thin strands with opposite $I_D$  values that over time form a loop of complex red- and blueshifts. At $t=83$~s, a substructure of negative $I_D$ at the right footpoint starts to form and travels up and toward the apex of the loop. It eventually starts to fade and loses its shape at $t=155$~s. At the same time instance, there are two more loop structures visible with the same Doppler-shifted patterns, negative $I_D$ on the left and positive $I_D$ on the right. One of these features appears already at $t=19$~s with predominantly positive $I_D$, until $t=102$~s when a negative layer of $I_D$ connects. The same feature is almost invisible at $\mu=1$. Only a small point where the H$\alpha$ Doppler shifts of opposite signs meet at $t=102$~s and $t=108$~s (Fig.~\ref{fig:1.1}) reveals its position. The second loop-like feature that appears at $\mu=1$ is also visible in the same snapshot at $\mu=0.66$ as the highest and largest loop emanating from the left footpoint. At this angle, at time instance $t=155$~s, it is the faintest looking loop in the Dopplergrams.\\
~\newline
The Si~IV emission at both footpoints is much better visible at $\mu=0.66$ to the $\mu=1$ case. There is a string of small bright points forming in the same region as the loop-like structure in the Dopplergram. The bright points form at first on the left and then to the right, until $t=83$~s when a large emission feature becomes well visible to the right of the apex. The structure grows in size and intensity, forming an arch above the H$\alpha$ loop. At $t=155$~s, the Si~IV emission is slightly faded but some traces of it remain.\\
\begin{figure}
   \centering
   \includegraphics[width=\linewidth,trim= 5cm 0.5cm 0cm 0cm,clip=true]{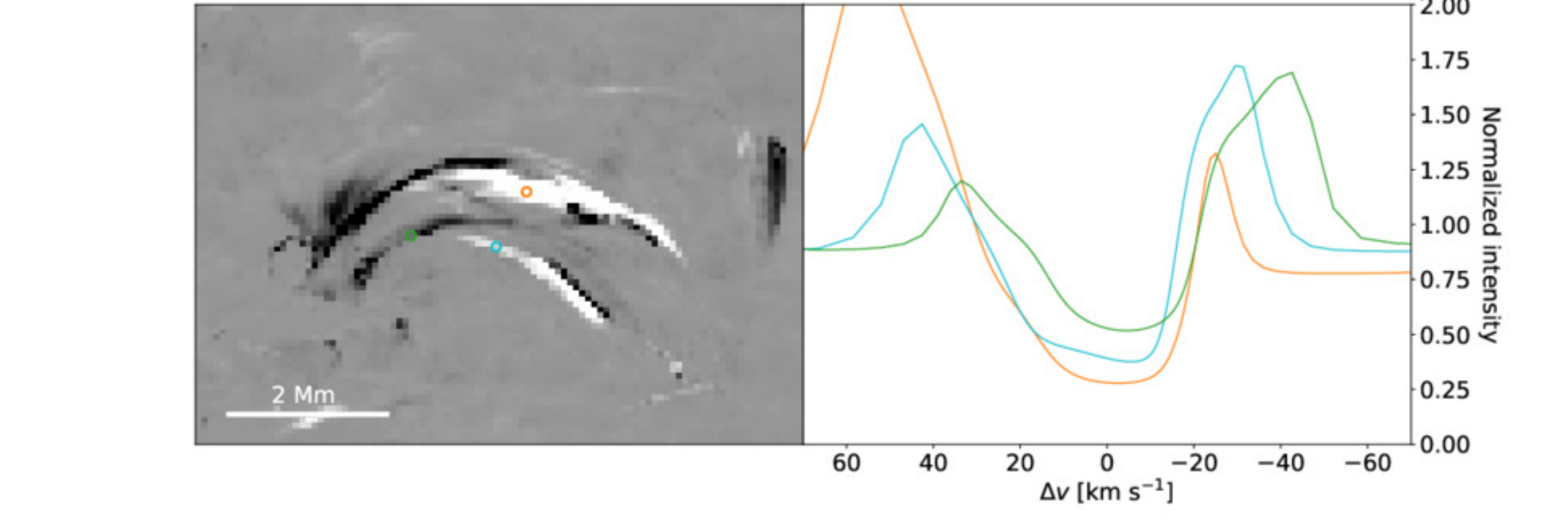}
   \caption{\textit{Left:} The loop in synthetic H$\alpha$ at $\mu=0.66$ at $t=102$ s, where three chosen pixels are indicated by circles in green, orange and blue. \textit{Right:} the normalised H$\alpha$ intensity for the pixels indicated with the respective colour in the left subplot. 
   }
    \label{fig:5.1}%
\end{figure}    
\begin{figure}
 \centering
\includegraphics[width=\linewidth,trim= 1cm 0.7cm 6.3cm 3.7cm,clip=true]{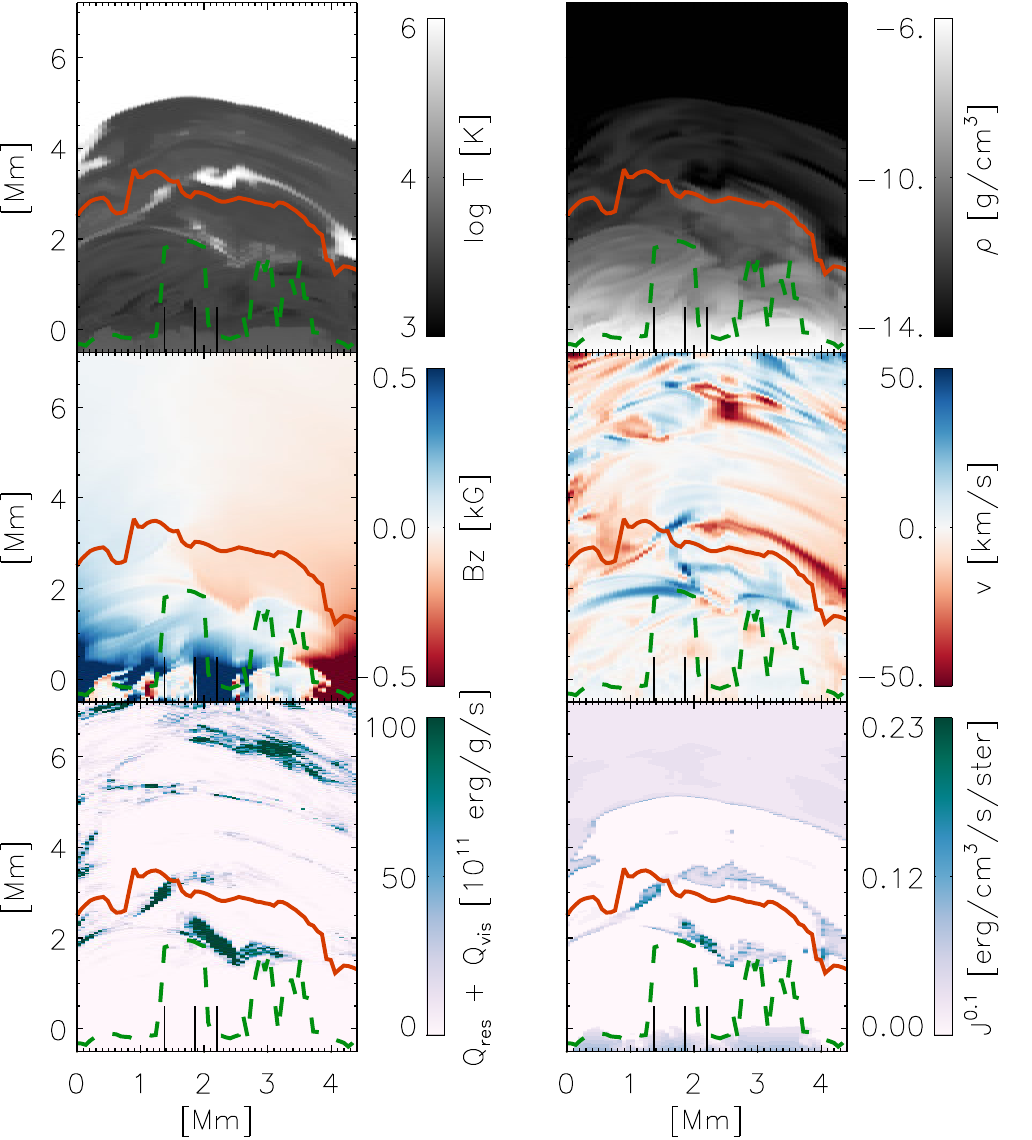} 
  \caption{Vertical cut along the dotted line plotted in Fig.~\ref{fig:3}. From top to bottom, left to right: temperature, density, vertical components of magnetic field and velocity, sum of resistive and viscous heating terms and total line emissivity for $1394$~\AA. Red line marks the formation height of H$\alpha$ line core. Green line marks the formation height of H$\alpha$ blue wing at $v_D = -53$ km s$^{-1}$. Black vertical lines, from left to right, mark the position of the red, green and white pixels in Fig.~\ref{fig:3} respectively. }
\label{cut}
\end{figure}
\newline
H$\alpha$ spectra of three pixels on the loop-like structure at $t=102$~s in $\mu=0.66$ are displayed in the right subplot of Fig.~\ref{fig:5.1}. The H$\alpha$ spectra in the pixel marked with a blue circle in the left subplot are represented by the blue curve in the plot to the right, and so on.  All three line profiles are asymmetric, similarly to the ones that appear in the case of the $\mu = 1$. The emission peaks in line wings are still there also at the slated view. The difference to the $\mu = 1$ case is the largest closer to the line centers where there is an indication of knees. The green and blue pixels that are on the opposite sides of one of the loop-like structures show the line core shifted in opposite directions.

\begin{figure*}
    \centering
    \includegraphics[width=180mm]{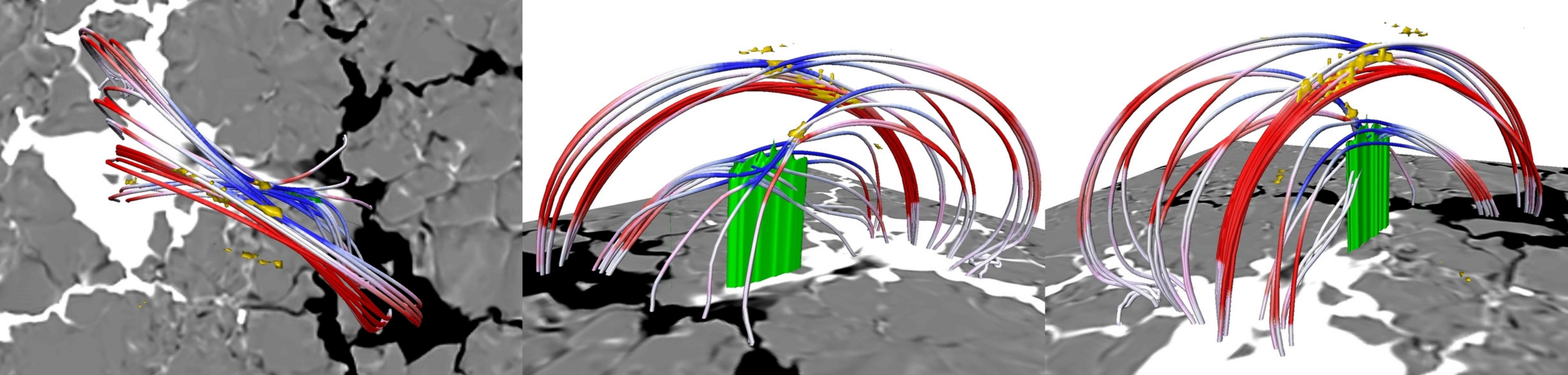}
    \caption{3D rendering of the area of interest over the photospheric vertical component of the magnetic field at $t=102$ s. Here, we choose to plot the formation height of the H$\alpha$ blue wing at $- 53$~km~s$^{-1}$ (green isosurface), the formation height of the Si~IV~1394~$\AA$ line (yellow isosurface) and magnetic field lines of the structure in blue and red, corresponding to up- and downflowing material with $v_z = \pm 122$~km~s$^{-1}$ respectively. (will be updated) The figure is produced with VAPOR \citep{atmos10090488}. }
    \label{fig:fh_vapor}
\end{figure*}
\begin{figure*}
 \centering
\includegraphics[width=\linewidth,trim= 0cm 0.8cm 0cm 0.4cm,clip=true]{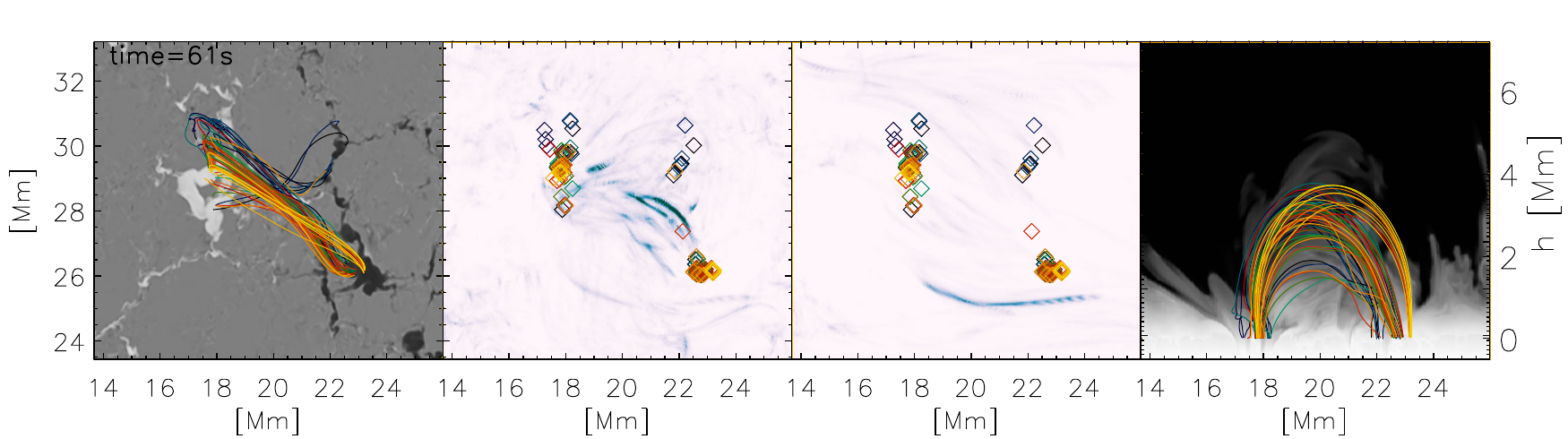} 
\includegraphics[width=\linewidth,trim= 0cm 0.8cm 0cm 0.4cm,clip=true]{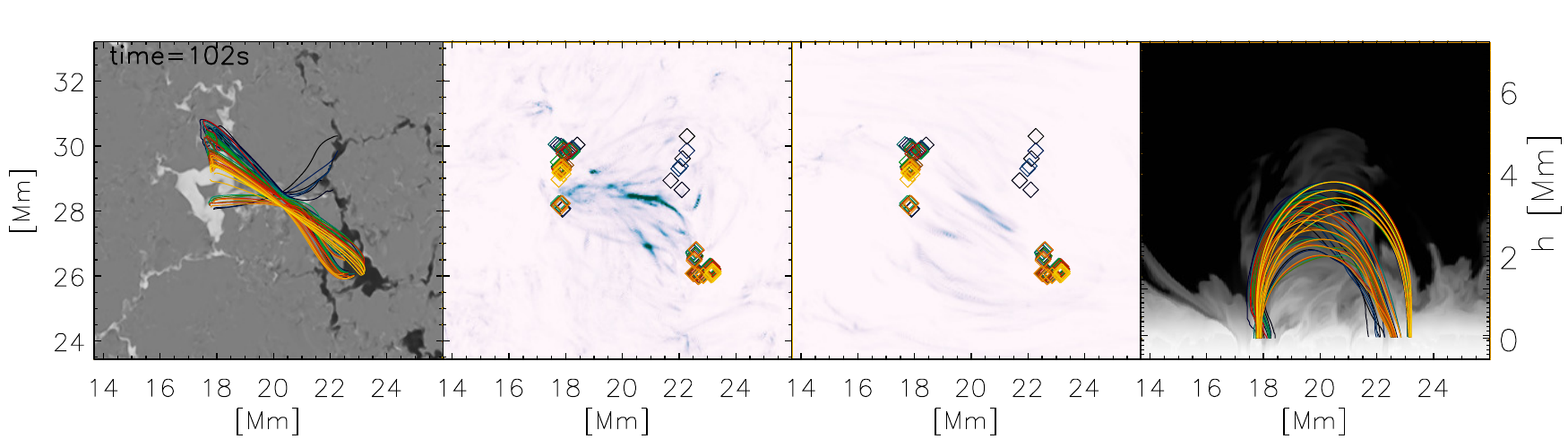} 
\includegraphics[width=\linewidth,trim= 0cm 0cm 0cm 0.4cm,clip=true]{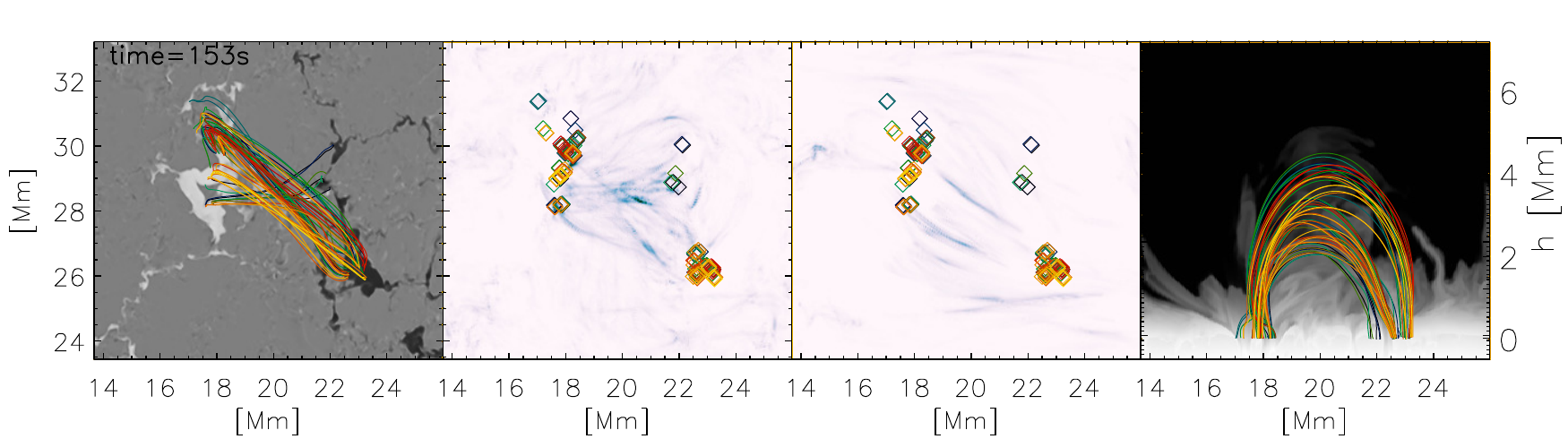} 
  \caption{Evolution of traced magnetic field lines. On the left, traced magnetic field lines are plotted over the vertical component of the photospheric magnetic field. In the middle, the position of the corresponding footpoints is marked over the a sum of resistive and viscous heating terms integrated over the height range $[0.6-3.8]$~Mm (second column) and $[2.7-7]$~Mm (third column). On the right, the projection of the traced field lines to the z-axis. The background shows density along the horizontal cut at y$=28$~Mm. The movie is available as online material.}
\label{line_trace}
\end{figure*}
\begin{figure}
 \centering
\includegraphics[width=\linewidth,trim= 0.5cm 0cm 5cm 0.6cm,clip=true]{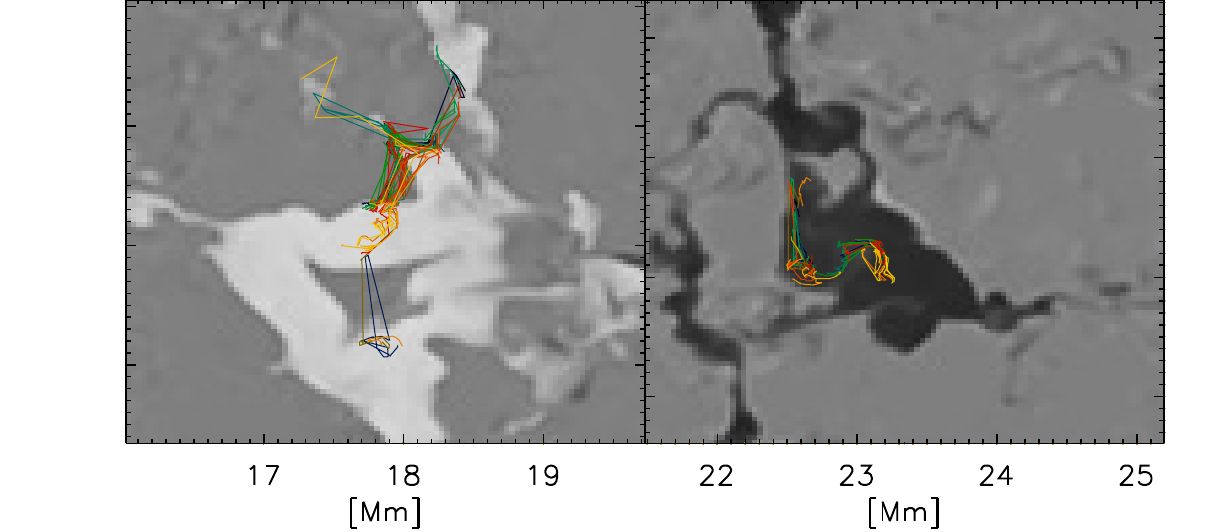} 
  \caption{Footpoints of the chosen traced field lines. The colors are the same as in Fig.~\ref{line_trace}.}
\label{footpoints}
\end{figure}
\subsection{Formation of the synthetic feature}
Figure~\ref{cut} shows a vertical cut along the dotted line marked in Fig.~\ref{fig:3}. The cut covers the bright feature visible in the H$\alpha$ blue wing and crosses the Si~IV emission feature. Figure~\ref{cut} shows various atmospheric parameters as a function of height, together with the emissivity $j_{1394}$ in the bottom right panel. Formation heights of both the H$\alpha$ blue wing and line core are plotted over all 2D cuts of different quantities. In the temperature map (upper left panel), two hot pockets are visible, of which the lower one lies at $1.5-2$~Mm height. This hot region generates the peaks visible in the source functions in Figs.~\ref{fig:4.1} and ~\ref{fig:4.2}. This is exactly where the H$\alpha$ blue wing forms, but also bright points visible in Si~IV images. The same region shows an oppositely directed magnetic field (middle left), an increase in the heating rate (bottom left) and strong upflows (middle right). At the apex of the upflowing feature, we can see a shape that resembles the inclined Y. All this coincides with a slight decrease in density. The density drop in the second hot pocket formed higher up at $\sim 3$~Mm, is greater. This region overlaps with strong downflows, adjacent upflows and increased heating rates. The H$\alpha$ line core is formed below this hot pocket which is also visible in the fourth row in Figs.~\ref{fig:1.1}. There, the line core intensity noticeable drops along a thin line where Si~IV emission increases ($t=102$~s) and space between H$\alpha$ line core filaments is formed. The resulting Si~IV line profiles formed in this region are shown in Fig.~\ref{fig:6.3} and ~\ref{fig:6.4}.\\
\newline
Figure~\ref{fig:fh_vapor} shows again the same snapshot at $t=102$~s, but now visualised in 3D. The magnetic field lines are traced in the regions where the H$\alpha$ blue wing and Si~IV~1394~$\AA$ lines are formed. Isosurfaces in yellow correspond to the largest values of j$_{1394}^{0.1}$ visible in Fig.~\ref{cut}. The green isosurface represents the formation height of the H$\alpha$ blue wing at $- 53$~km~s$^{-1}$. The vertical component of the photospheric magnetic field is displayed as a surface in black and white. The magnetic field lines are color-coded so that the blue and red colors indicate the up- and downflowing plasma, clipped at $\pm 122$~km~s$^{-1}$. The figure is divided into three subplots, where the right subplot shows the feature at $\mu=1$ and the other two at two slanted angles. The figures show two groups of yellow surfaces. The lower, smaller group lies just at the edge of the green isosurface, i.e.\ where the H$\alpha$ blue wing is formed. The magnetic field lines traced from this region show the same upflowing Y-shape visible in Fig.~\ref{cut}. The footpoints of these lines spread over a large area where negative polarity resides. The other footpoint is connected to the outer edge of the positive polarity. The field lines that are traced from the higher yellow region naturally rise further up. Their footpoints are shifted towards the centers of magnetic features. Their direction outlines the direction of the Si~IV emission feature in Fig.~\ref{fig:3}. These field lines show upflows at the apex and strong downflows all the way down to the photosphere. A local twist is visible in both sets of magnetic field lines.\\
\newline
To learn what drives the formation of these twists, we trace the magnetic field lines as curvatures in 3D space \citep{2015ApJ...802..136L}. The time cadence between snapshots used for the field tracing is around $2.5$~s. The starting seeds are planted across not along the H$\alpha$ blue-wing emission feature visible in Fig.~\ref{fig:3}. They are placed equidistantly at the height range from $2$ to $4$~Mm. Figure~\ref{line_trace} shows $60$ field lines at three time instances. The field topology at $t=102$~s is almost the same as in Fig.~\ref{fig:fh_vapor} which confirms that the starting seeds are well spread. Figure~\ref{line_trace} shows that the two opposite polarities connect in a complex way. Most of the field lines sampled with these seeds are connected to the outer edge of the positive polarity. Lying under them, a few loops extend in a different direction, forming an angle with the rest of the loops above. One of the footpoints of these low-lying loops ends in the extended branch of the negative polarity that undergoes a rapid movement downwards. As these footpoints move, the field lines rise and touch the ones higher up. An upward zip-like motion is visible in the accompanied movie, especially in the range at $x=20-22$~Mm, exactly where the H$\alpha$ features appear for the same time instances. The sampled field lines are most concentrated at the $t=102$~s which is expected since this is the time instance that is used to start line tracing. After that, the field lines unwind and spread out, some of which rise higher, close to $4$~Mm.\\ 
~\newline
The middle panels of Fig.~\ref{line_trace} show a sum of resistive and viscous heating terms integrated over two different height ranges. One is between  $[0.6-3.8]$~Mm and the other between  $[2.7-7]$~Mm to showcase features formed lower and higher up. The heating pattern in form of stripes that connect two polarities is visible in both middle columns. The highest heating rates appear where the stripes seem to cross. The heating happens on different temporal scales. There are very short, transient events on one side and long-lasting patterns on the other. The events in the latter group seem to outline exactly the features visible in the H$\alpha$ line. The heating in the higher layers outlines the Si~IV emission feature starting with a long fibrilar structure at $t=61$~s, visible also in the corresponding image in the third row of Fig.~\ref{fig:1.1}. \\
~\newline
Figure~\ref{footpoints} zooms into two main footpoints and the corresponding paths they cross during these few minutes. The figure shows footpoints of only fairly stable magnetic field lines. The positive footpoints are very unstable, displaying frequent jumps from one side to the other of the magnetic concentration. The side projected field lines in Fig.~\ref{line_trace} also show a lot of dynamics close to the positive end. The negative polarity shows a more linear behavior, with all footpoints undergoing the motion in the same direction, counterclockwise. The shape of the feature suggests that there might be an onset of a vortex flow, but the time interval is too short to be certain. As a result of both types of movement, the field lines get wind and unwind in a matter of minutes.

\section{Discussion}
\label{s:discussion}
\begin{figure}
\centering
\includegraphics[width=0.65\linewidth,trim= 0.1cm 0cm 0.3cm 0.1cm,clip=true]{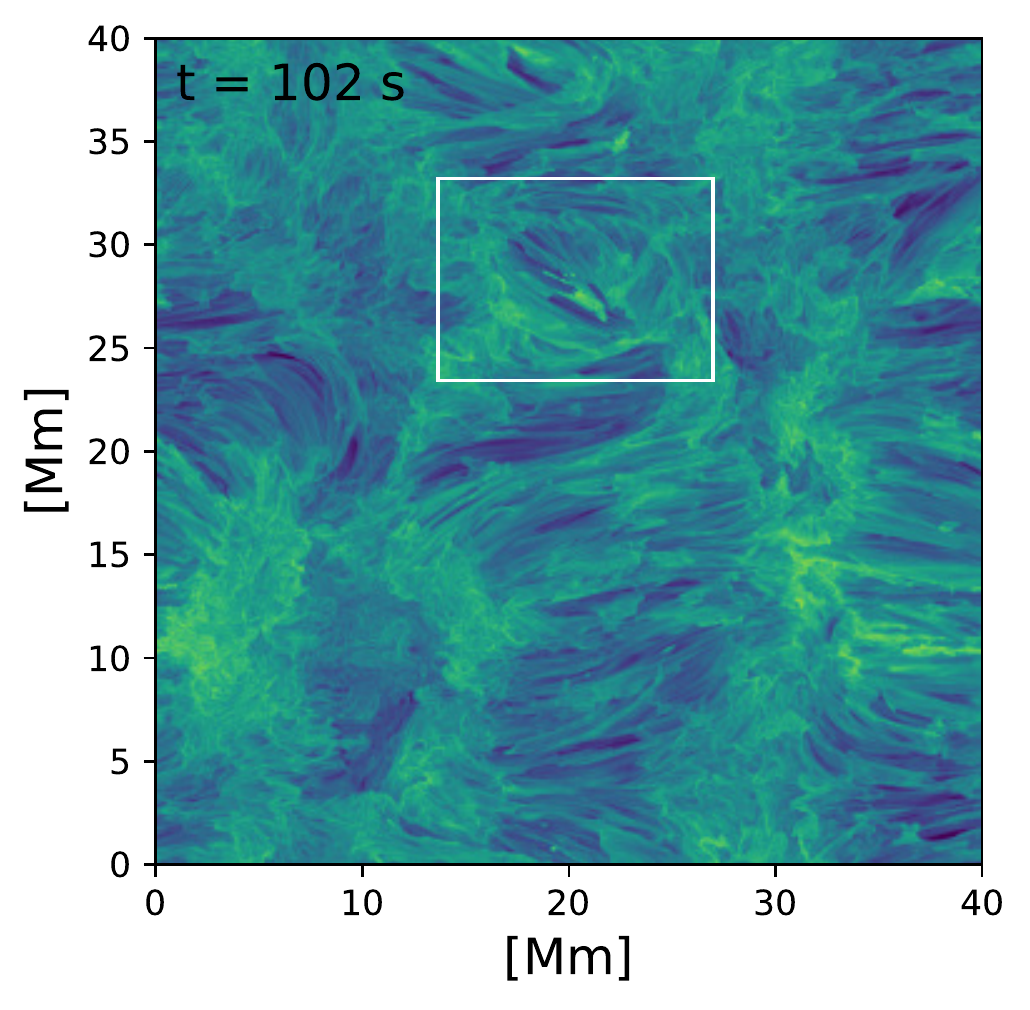} \\
\includegraphics[width=0.3\linewidth,trim= 1.5cm 1.4cm 0.2cm 0.2cm,clip=true]{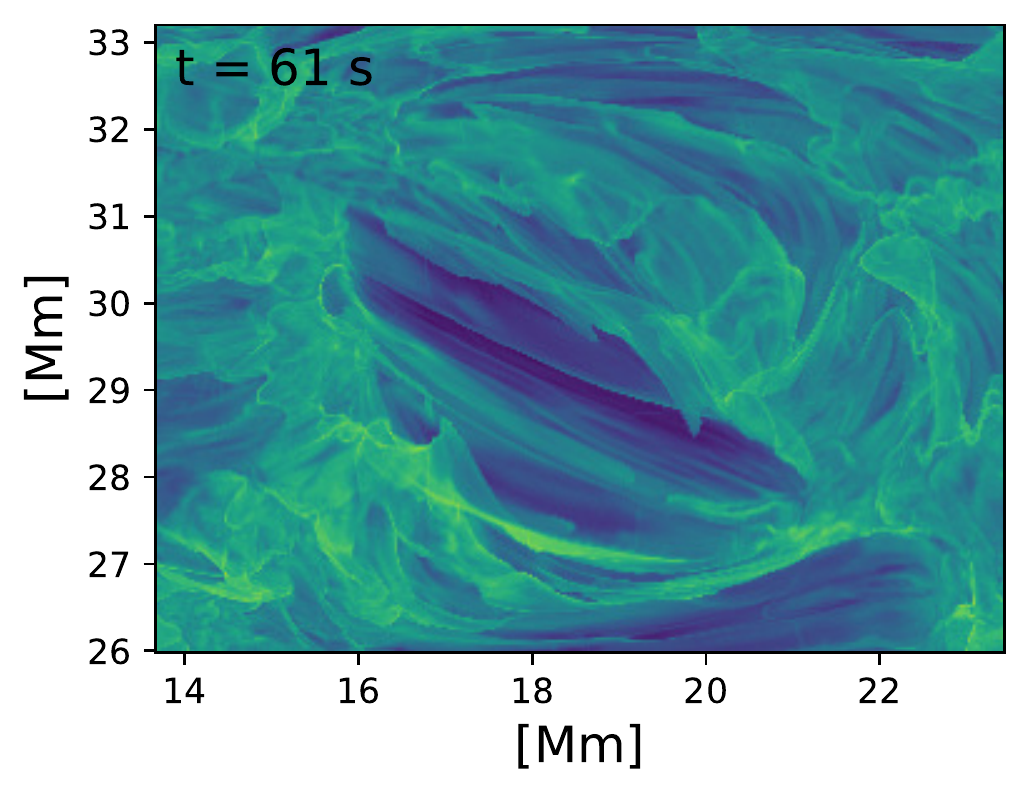}
\includegraphics[width=0.3\linewidth,trim= 1.5cm 1.4cm 0.2cm 0.2cm,clip=true]{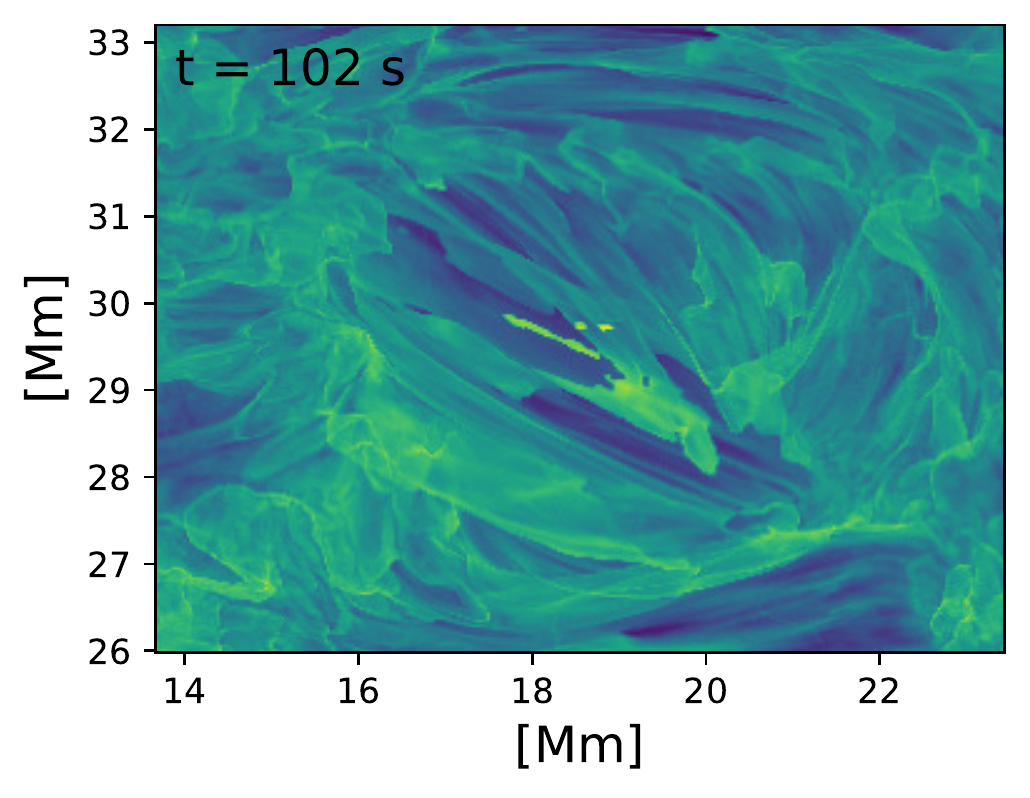}
\includegraphics[width=0.3\linewidth,trim= 1.5cm 1.4cm 0.2cm 0.2cm,clip=true]{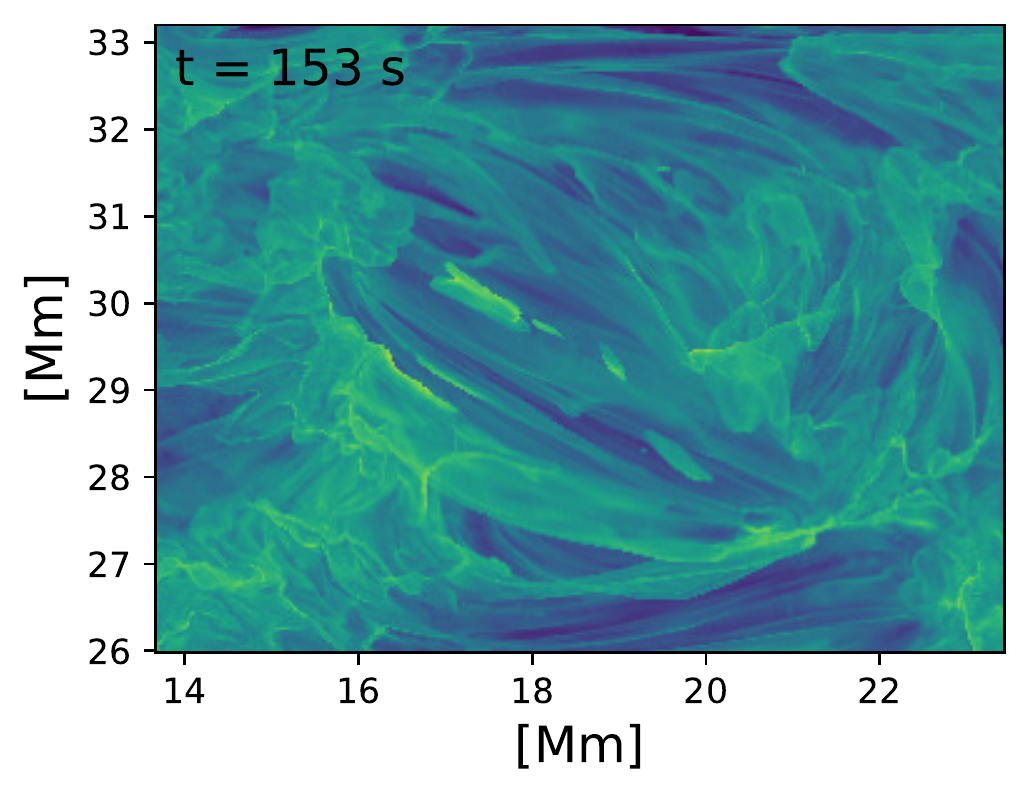}
  \caption{The emission measure for the lg($T$ /K)$ = [5.0 , 5.5]$ temperature interval. Bottom row shows zoom in of the area marked with the white rectangle. The area size and time steps are the same as in Fig.~\ref{line_trace}.}
\label{aia}
\end{figure}

We study an example of a bipolar system in rMHD simulations that shows the formation of several small-scale loop-like structures in synthetic chromospheric and TR observables. These structures appear in several places between the two opposite polarities over a few minutes we consider in this study. In the synthetic H$\alpha$ Dopplergrams we see a complex mix of strands of plasma with negative and positive values extending to both of the magnetic polarities. At the same time, we see signatures in synthetic Si~IV spectral lines. We find our results consistent with observations in terms of geometry, lifetime, dynamics and morphology of the features visible in Si~IV and the H$\alpha$ wings \citep{2018A&A...611L...6P,2016ApJ...826L..18B,2014Sci...346E.315H}. The half-length of the loops studied here is $ \approx 2$~Mm which is what is found in observations. Their lifetimes are also consistent with observations that exhibit distribution with an average of $107$~s \citep{2016ApJ...826L..18B}. One of the modeled loop-like structures first appears in the H$\alpha$ wings and then a minute later in Si~IV. This indicates that after approximately $1$~minute, the material is heated to transition region temperatures and the fraction of Si~IV becomes significant. Other loop-like structures display different development depending on the viewing angle.\\
\newline
\citet{2018A&A...611L...6P} reported a spatial overlap of H$\alpha$ and Si~IV which is only partly true in our case. There is some overlap of the loop-like structure in the H$\alpha$ wings with small bright points in Si~IV, but the more pronounced Si~IV emission loop is forming in a different spatial plane. Only due to the projection effects it appears like the Si~IV emission seemingly sits on top of the structure in the H$\alpha$ although the two features form more than $1$~Mm apart. We also find a case when signatures in H$\alpha$ are situated above the Si~IV emission in the slanted view. Detailed examinations reveal, however, that two types of signatures correspond to different magnetic field strands.  This urges caution in interpreting observations. \\
\newline
There are, nevertheless, locations where H$\alpha$ and Si~IV signatures coincide. The model thus supports the interpretations given in observational studies \citep{2016ApJ...824...96T} which state that in some cases the hot pockets can produce signatures in both observables co-spatially. The model shows similarities with features like Ellerman bombs, UV bursts and Explosive events \citep{2018SSRv..214..120Y}. The last type of features, as interpreted by \citet{2017MNRAS.464.1753H}, can have different morphologies and undergo dissimilar evolution. We find events in our model that reproduce these observations rather well. \\
\newline
H$\alpha$ profiles produced by the model are asymmetric with pronounced emission in the line wings, typical Ellerman bomb or a “mustaches” type of spectra \citep{2013A&A...557A.102B,2016A&A...590A.124R}. They are formed when the increase in temperature and density is accompanied by high velocities \citep{2017A&A...601A.122D}. The emission peak intensity in the line wings does not correlate with the formation height but more with the summit in the source function, as the four-panel diagrams illustrate. Also, the asymmetry in the emission peaks seems not to be necessarily correlated with the velocity in the regions where the line core forms, as suggested first by \citet{1983SoPh...87..135K}. \\
\newline
However, several discrepancies can be found if we compare the synthetic H$\alpha$ profiles with observations by \citet{2018A&A...611L...6P}. First, observations do not show any H$\alpha$ profiles with wings in emission. It is possible that some wing emission occurs outside the observed wavelength range, so it is not observed. It is also clear from our analysis that higher velocities would shift the emission peaks further away from the nominal line center. It is also likely that our model underestimates the velocities and overestimates the size of the region heated to the temperatures that H$\alpha$ is formed at. Second, we do not see  inverted Y-shape in the synthetic dopplergrams. The shape is, however, visible in the model, in the field configuration as demonstrated in  Fig.~\ref{cut} and it is not a product of emerging flux reconnecting with the overlying field. It may be that we do not sample the event from the right angle so it gets its imprint on  H$\alpha$ spectra. Third, only one observed example resembles the pattern we see in our synthetic dopplergrams. Namely, the Doppler shifts with opposite signs on opposite parts of the loop are visible only in case A in \citet{2018A&A...611L...6P}. Cases B and C show only one sign Doppler shift. It is not clear from the paper which case of the two is more frequent and why this might happen. Looking into our results, especially the long loop visible at $t=155$~s, the Doppler pattern depends very much on the viewing angle.\\
 %\sout{Both of the wings display a "knee" which is usually seen in a chromospheric plasma which is experiencing a strong Doppler shift. Indeed, the intensities at the knees are formed in regions that experience steep velocity gradients. This type of knee can also be seen in the profiles of the observations done by \citet{2018A&A...611L...6P}.} 
\newline
To synthesise Si~IV lines we used the full 3D non-LTE calculation instead of coronal approximation. As shown by \citep{2015ApJ...811...80R} on example of C II line, this treatment leads to lower formation temperatures and higher intensities (Fig.~9 in that paper). 
The synthetic Si~IV emission is concentrated above and around  magnetic features. The increased Si~IV intensity is even more visible in the slighted view, resembling facular brightening. The same is reported observationally \citep{2015ApJ...814...70R,2021ApJ...916...36A}.\\
\newline
The intensity of the synthetic Si~IV 1394 $\AA$ and 1403 $\AA$ emission lines are comparable to the ones observed in local heating events like UV bursts \citep{2018SSRv..214..120Y}, but not as strong as the ones we should expect to find in flare ribbons \citep{2019ApJ...871...23K}. We find a range of synthetic emission profiles, both symmetric and asymmetric, blue- or red-shifted. The dopplershift are of the order of $50$~km~s$^{-1}$. As the Figs.~\ref{fig:6.1}-~\ref{fig:6.4} show, the bulk of Si~IV emission comes from small hot pockets with temperatures of $10^{4.9}$~K. When these temperatures coincide with electron densities larger than $2 \times 10^{11}$~cm$^{-3}$, the opacity effects become considerable and the lines optically thick. In these cases we find that the peak ratio of Si~IV 1394 $\AA$ and 1403 $\AA$ drops to $\approx 1.6$. This is in agreement with other studies which state that the intensity ratio reflects the ratio of the source functions instead \citep{2015ApJ...811...80R,2019ApJ...871...23K} and can have any value in the optically thick regime. The wavelength integrated ratio in these cases is in $1.6-1.8$ range. \cite{2020ApJ...894..128T} found the same ratio in the quiet Sun areas as well as periphery of emerging regions. As the event studied here is smaller in size and energy, it fits to these observations quantitatively.\\    
\newline
In this paper, we have not considered effects from \textit{non-equilibrium ionisation} (NEQ). The effect is important for both species H and Si \citep{2007A&A...473..625L,2015ApJ...802....5O} when the timescales of the dynamics are shorter
than the typical ionization and recombination times of the emitting atoms/ions. When NEQ are taken into account, \cite{2015ApJ...802....5O} show that Si IV can originate at temperature as low as $10^{4}$~K and the resulting intensity can be from $10\%$ to $40\%$ higher depending on the region. Given that the process is slow and the feature that we study here is relatively long lived, we do not think that the difference coming from NEQ would lead to qualitatively different result. However, using the generalized Ohm's law including the effects of ambipolar and Hall drift \citep{2012ApJ...750....6C,2012ApJ...753..161M} might potentially change the area of interest. This is also not certain given the recent study that shows that differences including both effects together are not fundamental \citep{2020A&A...633A..66N}.\\ 
\newline
Tracing the field lines shows that there are three types of footpoint drivings that lead to the observable dynamics. The main observational features are product of a reconnection which takes place as the footpoints of low-lying field lines move fast towards the main bipolar system. At the same time, the two opposites polarities where the main bipolar system is rooted undergo different type of dynamics. The positive polarity shows a complex flux concentration where field lines constantly reconnect and as a result the H$\alpha$ dopplergrams show faster and more frequent flows. The negative footpoint shows an indication of a vortex flow formation. All these types of  motions lead to stresses in the magnetic field of structures connecting opposite polarities. The injected stresses lead to gradients in the field that cause current sheets to form, leading to heating and dissipation to occur at multiple places over the period of a few minutes. This is agreement with observations that show recurrence and reappearance of the loop-like structures visible in Si~IV slit-jaw movies \citep{2014Sci...346E.315H}. We expect that these features are sustained by complex footpoint motions similar to the ones seen in our model.  \\
\newline
Evolution maps of averaged resistive and viscous heating terms show many places where heating signatures have morphology visible in observables. The synthetic Si~IV loops show no twisted structures in agreement with \citep{2022ApJ...933..153P}, but the field line tracking reveals twist in the apex of the loop-like shape where the Si~IV is being formed which is in agreement with \citet{2015ApJ...811..106H}. The loops observed in \citet{2013A&A...556A.104P} and \citet{2017A&A...599A.137B} have similar properties as the one that we see in our results, in terms of geometry and lifetime and even temperature as illustrated in Fig.~\ref{aia}. The figure shows the emission measure for the lg($T$ /K)$ = [5.0 , 5.5]$ which should be the same temperature interval that both EUI~173 passband and AIA 171 channel are sensitive to. The HI-C 193 images should trace higher temperatures than this, but this is with assumption that emission comes from corona and that coronal approximation is valid. Many of the features produced by our model is situated in chromosphere. \\
\newline
% compare to campfires
Figure~\ref{aia} shows several features that are also partly visible in Si~IV lines: a loop at $t=61$~s in the lower part of ROI, and the two counterparts of the Si~IV emission features in between opposite polarities at $t=102$~s  and $t=153$~s.  The field configuration around some of the heating features that appear in the model is similar to one of the four possible scenarios of the campfire configurations listed in \citet{2021A&A...656L...4B}. This is also what \citep{2021A&A...656L...7C} find in their rMHD model. The two models, presented here and in \citep{2021A&A...656L...7C} differ in the modeled mean field strength. So the former represents more active regions on the Sun and models network as opposed to the latter which contains only a weak network. Possibly mainly as a result of more flux, the chromosphere is lifted higher up in our model than in the latter. As a result, the relevant features form at similar heights as in  \citep{2021A&A...656L...7C} but at a dex higher electron densities.\\    
\newline
A study of adjacent photospheric magnetic features and their evolution by \citet{2022A&A...660A.143K}, indicate that a large majority of campfires appear within a bipolar system. They do state that there is no significant movement of the two opposite polarities in half of the cases, but still, the authors interpret the appearance of the campfire as a sign of reconnection. They also find that a quarter of all analyzed campfires occur 'within a region with randomly scattered small magnetic features' in a very close vicinity. A closer inspection of the representative example of this category of campfires reveals that there are, in fact, larger magnetic elements a few Mm away from the feature identified as the campfire. All this resembles the ROI from our model. The observational facts related to the campfires do not exclude that the appearance of these features is a product of the evolution similar to the magnetic system in our model where features appear in one of the heated regions at various heights as a result of rapid and complex footpoint motion.

\section{Conclusion}
\label{s:conclusions}
%The new theoretical models offer new insights into the solar atmospheric dynamics and heating processes. Our theoretical concepts and interpretation of observations are being alliterated as a result of it. For example, concept of thin flux tube is being replaced with “coronal veil” hypothesis which explains that the volume counterparts for 'coronal loops' are in fact extremely inhomogeneous and complex in shape \citep{2022ApJ...927....1M,2022A&A...658A..45B}. At the same time, the new instrumentation is providing observations of the solar atmosphere in increasing finer detail and further challenging the models. 
%Similar signatures of small-scale reconnection are found in low-lying cooler active- region loops that only reach transition-region temperatures. These signatures are visible in sub-arcsecond resolution IRIS Si IV spectra as strong brightenings with anomalously broadened profiles and are associated with the formation of loops (Huang et al., 2015, 2017).
Our model largely reproduces the properties of loop-like features simultaneously observed in H$\alpha$ and Si~IV. Although the model has some limitations it illustrates the following:
\begin{itemize}
    \item Two magnetic concentrations of opposite polarity $\approx 8$~Mm away from each other are connected by numerous thread-like structures in the viscous and resistive heating, the majority of which are situated in the chromosphere, and some in the transition region and corona.
    \item Convective motions and complexity of magnetic concentration combined lead to a generation of many localised recurrent heating events at different heights and temperature regimes, including in the relatively dense chromosphere.
     \item Magnetic reconnection in the simulated loop system can take place without flux emergence.
     \item The features observed in  H$\alpha$  and Si~IV highlight the location of the current sheets.
    \item The appearance of the H$\alpha$ Dopplergrams depends very much on the viewing angle.
    \item Projection effect make H$\alpha$ and Si~IV features appear cospatial, even if their are occurring at different locations in the atmosphere.
    %Features related to different magnetic threads can seem connected at the slanted viewing angle.
\end{itemize}
The model is not necessarily “correct”, but rather
gives guidance and insight for interpreting the observation and improving our understanding of these small-scale phenomena and their place in the solar atmosphere.

\begin{acknowledgements}
This project has received funding from Swedish Research Council (2021-05613), Swedish National Space Agency (2021-00116) and the Knut and Alice Wallenberg Foundation. This research data leading to the results obtained has been supported by SOLARNET project that has received funding from the European Union’s Horizon 2020 research and innovation programme under grant agreement no 824135. This material is based upon work supported by the National Center for Atmospheric Research, which is a major facility sponsored by the National Science Foundation under Cooperative Agreement No. 1852977. The calculations were performed on resources provided by the Swedish National Infrastructure for Computing (SNIC) at the National Supercomputer Centre (NSC) at Linköping University and the PDC Centre for High Performance Computing (PDC-HPC) at
the Royal Institute of Technology in Stockholm. 
\end{acknowledgements}


\bibliography{Synthetic_UFS.bbl}{}
\bibliographystyle{aa.bst}

\end{document}